\documentclass[preprint,12pt]{elsarticle}

\usepackage{amssymb}
\usepackage{amsmath}
\usepackage{multirow}
\usepackage{subcaption}
\usepackage{microtype}
\usepackage{setspace}
\usepackage{makecell}
\usepackage{BOONDOX-cal}
\usepackage{xcolor}
\usepackage{colortbl}
\usepackage{caption}
\usepackage{geometry}
\journal{Signal Processing}

\begin{document}
\begin{sloppypar}
\begin{frontmatter}



\title{Modulation Feature Enhancement with a Multi-Stage Attention Network for Underwater Acoustic Target Recognition}


\author[1,2]{Jiaping Yu} %% Author name
\author[1,2]{Shefeng Yan\corref{cor1}}
\ead{sfyan@ieee.org}
\author[1]{Linlin Mao}
\author[3]{Zeping Sui}
\author[1,2]{Chunjin Jiang}
\cortext[cor1]{Corresponding author}
\affiliation[1]{organization={Institute of Acoustics, Chinese Academy of Sciences},%Department and Organization 
            city={Beijing},
            postcode={100190}, 
            country={China}}
\affiliation[2]{organization={University of Chinese Academy of Sciences},%Department and Organization 
            city={Beijing},
            postcode={100049}, 
            country={China}} 
\affiliation[3]{organization={School of Computer Science and Electronics Engineering, University of Essex},
			city={Colchester},
			postcode={CO4 3SQ},
			country={UK}
}                   

%% Abstract
\begin{abstract}

Underwater acoustic target recognition is critical for maritime applications, yet it faces challenges arising from the complex and diverse nature of ship-radiated noise. To address these issues, we propose a robust deep learning-based framework. First, we introduce a feature extraction and fusion method based on variational mode decomposition (VMD) and the 3/2-D spectrum to generate high-fidelity 2-D DEMON spectral features, which effectively capture modulation envelope information. To further enhance feature representation, we design a one-dimensional convolutional neural network (1-D CNN) integrated with a novel Multi-Stage Multi-Type Attention Mechanism (MMATT) that adaptively refines features at different network depths. Within this mechanism, we propose a Residual Channel-Independent Spectral Attention Mechanism (R-CISAM) and a Multi-Scale Separate-and-Fuse Spectral Attention Mechanism (MS-SFSAM). Moreover, to mitigate performance degradation caused by severe class imbalance inherent in real-world ship-radiated noise data, we devise an Adjustable Class-Balanced Focal Loss (ACBFL), which provides flexibility across tasks with varying degrees of imbalance. Experimental results on a real-world ship-radiated noise dataset demonstrate that the proposed solutions effectively enhance underwater acoustic target recognition performance.

\end{abstract}

\begin{graphicalabstract}
\begin{figure}
  \centering
  \includegraphics[width=\linewidth]{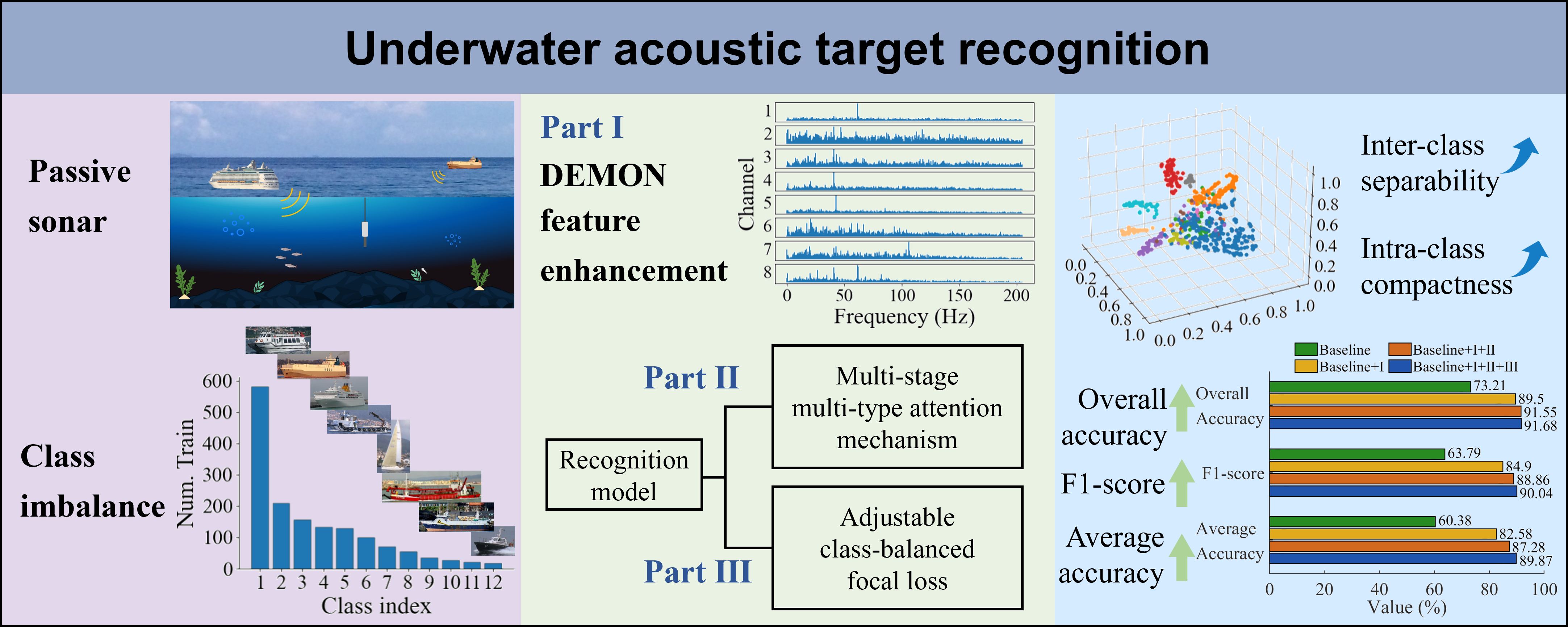}
\end{figure}
\end{graphicalabstract}

\begin{highlights}
\item A novel 2-D DEMON feature generated based on VMD and the 3/2-D spectrum is effective.
\item Multi-Stage Multi-Type Attention Mechanism enhances recognition performance. 
\item Two novel attention mechanisms improve accuracy and robustness over existing methods.
\item Adjustable Class-Balanced Focal Loss alleviates negative effects of data imbalance. 
\end{highlights}

%% Keywords

\end{frontmatter}



\section{Introduction}
\label{sec1}
Underwater acoustic target recognition constitutes a critical component of underwater acoustic signal processing. However, it is challenging due to the complexity of the marine environment and underwater acoustic propagation \cite{ref1, ref2, ref32}. Moreover, the scarcity of real-world data (stemming from target confidentiality and high acquisition costs) further exacerbates the difficulty of recognition.

Owing to the ability to process nonlinear, high-dimensional, and large-scale data, along with powerful feature learning capabilities, deep learning techniques have attracted intensive attention in underwater acoustic target recognition. Among these techniques, the convolutional neural network (CNN) is one of the most commonly used architectures, known for its good performance in capturing local features of input data \cite{ref3}, and has been widely applied to underwater acoustic target recognition \cite{ref4, ref5}.

The features input into the recognition network also have a significant impact on model performance. Ship-radiated noise primarily consists of three components: mechanical noise, propeller noise, and hydrodynamic noise. Modulation is a key characteristic of ship-radiated noise, with propeller noise being the primary source of modulating signals that contain crucial target features such as shaft frequency and blade number. Detection of envelope modulation on noise (DEMON) analysis extracts the low-frequency envelope spectrum by demodulating the time-domain signal. Pollara \textit{et al.} \cite{ref6} analyzed the DEMON spectra of a variety of small vessels and further revealed modulation inhomogeneity, concluding that the position, number, and relative amplitude of spectral peaks can be used to characterize vessel type.

The modulating signals in ship-radiated noise exhibit different modulation strengths across frequency bands. Conventional DEMON analysis, which performs demodulation on the full-band signal, can introduce significant errors. To address this issue, some researchers have proposed sub-band DEMON analysis methods based on band-pass filter banks \cite{ref7, ref8}. A critical yet challenging step in these methods is frequency band division, which is often simplified using uniform segmentation, a straightforward but coarse approach that lacks accuracy, robustness, and flexibility. Variational mode decomposition (VMD) \cite{ref9} is a completely non-recursive adaptive decomposition algorithm which optimally decomposes the input signal into a set of modes that all satisfy a specific sparsity property. VMD exhibits strong robustness, and the form of the intrinsic mode functions (IMFs) is well-suited for DEMON analysis. On the other hand, high-order statistics have been applied in underwater acoustic signal analysis due to their ability to suppress additive Gaussian noise and preserve signal phase information \cite{ref10, ref11}. Among these methods, 3/2-D spectrum requires less computation, making it suitable for practical applications. Moreover, it can eliminate non-phase-coupled harmonic components, which is beneficial for DEMON analysis \cite{ref12, ref13}. Nevertheless, few studies have applied VMD and 3/2-D spectrum for DEMON feature extraction within deep learning-based recognition methods.

To enhance model performance, numerous research efforts have focused on selecting critical information. Attention mechanisms, inspired by human cognitive attention allocation, are designed to focus on crucial information while suppressing irrelevant information \cite{ref14}. In underwater acoustic target recognition, Zhao \textit{et al.} \cite{ref15} incorporated a squeeze-excitation (SE) block and attentive statistics pooling into the recognition network. In \cite{ref16}, a convolutional block attention module was coupled with a multibranch backbone network, using time-frequency maps as input features. Yang \textit{et al.} \cite{ref17} adopted ResNet as the recognition network, which includes a frequency attention block and a comprehensive attention block that operates sequentially across channel, frequency, and time dimensions. However, most of these attention mechanisms are applied at a single position within the network, or the same mechanism is reused at multiple positions without variation.

However, features extracted at different network depths possess distinct characteristics and information content. Moreover, existing attention mechanisms are not sufficiently suitable for DEMON spectral features. The spatial attention mechanism (SAM) within the convolutional block attention module (CBAM) \cite{ref18} is a classic method widely used in computer vision and can be extended to the frequency dimension of spectral features. In the traditional SAM, the spatial attention weights are identical across different channels and are computed by aggregating information from all channels of the current feature map. For image recognition, different channels of middle-layer feature maps in CNNs represent diverse characteristics (e.g., edges, textures, colors), with the spatial positions of important information remaining consistent across these characteristics. However, this approach is unsuitable for features where key information is distributed differently across channels, such as the DEMON spectral features.

Deep learning is a data-driven approach; however, real-world ship-radiated noise data suffer from not only a small sample size but also class imbalance. Specifically, some categories occur much more frequently than others, leading to a long-tailed distribution in ship-radiated noise datasets. Training on such long-tailed data can produce biased recognition models that overfit head classes and underfit tail classes. Common strategies to address this issue include resampling \cite{ref19} and cost-sensitive loss \cite{ref20, ref21}. Typically, sample weights or resampling probabilities are set to be inversely proportional to class frequencies \cite{ref22}. However, some research \cite{ref23} has shown that this method is not always effective and may lead to poor performance. Consequently, an alternative weighting strategy proportional to the inverse square root of class frequencies was proposed \cite{ref24}. To address the hard-easy sample imbalance in dense object detection, Lin \textit{et al.} \cite{ref25} proposed focal loss. The ship-radiated noise recognition task may also encounter hard-easy sample imbalance due to variations in categories, operating conditions, and environments.

However, the class imbalance in ship-radiated noise recognition is less severe than in the original application of focal loss, and focal loss lacks the flexibility to generalize effectively to other tasks. Despite the significant imbalance in real-world ship-radiated noise data, few studies have addressed it. Dong \textit{et al.} \cite{ref26} proposed an exponentially weighted cross-entropy loss function, where an exponential function of the prediction probability is incorporated as a weighting factor into the cross-entropy loss function. Ma \textit{et al.} \cite{ref27} proposed a weighted cross-entropy function based on a trigonometric function. However, the calculation of weighting factors cannot be flexibly utilized for realistic imbalanced data.

\begin{table}[htbp]
  % \begin{center}
  \caption{Our contributions compared to the existing literature}
  \centering
  \fontsize{8}{12}\selectfont
  % \footnotesize
  \scalebox{0.99}{
    \begin{tabular}{ccccccccc}
      \hline
      \multicolumn{2}{c}{\textbf{Contributions}} & \textbf{This paper} & \textbf{\cite{ref12}} & \textbf{\cite{ref13}} & \textbf{\cite{ref15}} & \textbf{\cite{ref16}}  & \textbf{\cite{ref26}}  & \textbf{\cite{ref27}} \\
      \hline
      \multirow{3}{*}{\makecell[c]{\text{Feature} \\ \text{Extraction}}} & \text{VMD} & \text{\checkmark} & \text{} & \text{\checkmark} & \text{} & \text{}  & \text{}  & \text{} \\     
      & \text{3/2-D spectrum} & \text{\checkmark} & \text{\checkmark} & \text{\checkmark} & \text{} & \text{}  & \text{}  & \text{} \\     
      & \text{Fusion} & \text{\checkmark} & \text{} & \text{} & \text{\checkmark} & \text{\checkmark}  & \text{}  & \text{\checkmark} \\ 
      \hline
      \multirow{3}{*}{\makecell[c]{\text{Attention} \\ \text{Mechanism}}} & \text{SE Block} & \text{\checkmark} & \text{} & \text{} & \text{\checkmark} & \text{\checkmark}  & \text{}  & \text{} \\ 
      & \text{Frequency Attention} & \text{\checkmark} & \text{} & \text{} & \text{\checkmark} & \text{\checkmark}  & \text{}  & \text{} \\ 
      & \text{R-CISAM \& MS-SFSAM} & \text{\checkmark} & \text{} & \text{} & \text{} & \text{}  & \text{}  & \text{} \\
      \hline
      \multirow{3}{*}{\makecell[c]{\text{Loss} \\ \text{Function}}} & \text{Static weighting} & \text{\checkmark} & \text{} & \text{} & \text{} & \text{}  & \text{}  & \text{} \\ 
      & \text{Dynamic weighting} & \text{\checkmark} & \text{} & \text{} & \text{} & \text{}  & \text{\checkmark}  & \text{\checkmark} \\ 
      & \text{Adjustable Parameter} & \text{\checkmark} & \text{} & \text{} & \text{} & \text{}  & \text{}  & \text{} \\
      \hline
      \text{Experiment} & \text{Recognition} & \text{\checkmark} & \text{} & \text{} & \text{\checkmark} & \text{\checkmark}  & \text{\checkmark}  & \text{\checkmark} \\
      \hline
    \end{tabular}
  % \end{center}
  }  
  \label{table1}
\end{table}

Against this backdrop, we propose a two-dimensional DEMON spectral feature extraction and fusion method based on signal decomposition and the 3/2-D spectrum, along with a Multi-Stage Multi-Type Attention Mechanism (MMATT), and an Adjustable Class-Balanced Focal Loss (ACBFL). The main contributions of this study are summarized in Table \ref{table1}, which are detailed as follows:	

\begin{itemize}
\item We propose a 2-D DEMON spectral feature extraction and fusion method based on VMD and the 3/2-D spectrum to enhance the representation of modulation information in ship-radiated noise.

\item We develop an underwater acoustic target recognition network that integrates a 1-D CNN with our proposed MMATT, which adopts various types of attention modules at different network depths. 

\item Building upon our MMATT, we propose two attention mechanisms specifically designed for distinct feature characteristics: 1) a Residual Channel-Independent Spectral Attention Mechanism (R-CISAM) and 2) a Multi-Scale Separate-and-Fuse Spectral Attention Mechanism (MS-SFSAM), enabling more precise and effective information filtering than conventional approaches. 

\item To address the class imbalance in ship-radiated noise data, we design ACBFL, which effectively alleviates the adverse effects of long-tailed data distribution while offering flexible adaptability to tasks with varying degrees of imbalance.

\item Experimental results on the real-world dataset ShipsEar \cite{ref28} demonstrate that the proposed methods effectively improve recognition performance.
\end{itemize}

The remainder of this paper is organized as follows: Section \ref{sec2} introduces our underwater acoustic target recognition model, and details the proposed algorithms. Section \ref{sec3} describes the experiments and parameter settings. Section \ref{sec4} presents and analyzes the experimental results. Main conclusions of this paper are summarized in Section \ref{sec5}. 

\textit{Notations:} $\lVert \cdot \rVert_2$ denotes the $\ell_2$ norm, $\delta$ represents the Dirac distribution, $*$ denotes convolution, and $\mathbf{1}_{\left\{\cdot\right\}}$ represents the indicator function.

\section{Proposed Target Recognition Scheme}
\label{sec2}
%% Labels are used to cross-reference an item using \ref command.
%% Section text. See Subsection \ref{subsec1}.
\subsection{Feature extraction and fusion}
\label{subsec1}
\subsubsection{Variational Mode Decomposition}
%% Use \subsubsection, \paragraph, \subparagraph commands to 
%% start 3rd, 4th and 5th level sections.
%% Refer following link for more details.
%% https://en.wikibooks.org/wiki/LaTeX/Document_Structure#Sectioning_commands
The IMF of the decomposed signal in VMD is defined as an amplitude-modulated-frequency-modulated (AM-FM) signal, yielding
\begin{equation}\label{eq1}
u_{k}(t) = A_{k}(t) \mathrm{cos}\left[\phi_{k}(t)\right]
\end{equation}
where $A_{k}\left(t\right)$ and $\phi_{k}\left(t\right)$ denote envelope and phase, respectively. The VMD algorithm utilizes bandwidth as the sparsity prior of each mode, whereby the constrained variational problem is exploited to obtain the estimated IMFs. The constrained variational problem can be formulated as
\begin{equation}\label{eq2}
\underset{\left\{u_{k}\right\},\left\{\omega_{k}\right\}}{\mathrm{min}}\left\{
\sum_{k = 1}^{K}\left\|\mathrm{\partial}_{t}\left[(\delta(t) +\frac{j}{\pi t}) * u_{k}(t)\right]e^{-j\omega_{k}t}\right\|_{2}^{2}\right\}, \quad \mathrm{s.t.} \sum_{k = 1}^{K} u_{k} = s
\end{equation}
where $s$ represents the original signal, $\left\{u_k \right\}|_{k=1}^K$ and $\left\{\omega_k \right\}|_{k=1}^K$ denote the IMFs and the corresponding center frequencies, respectively. VMD can adaptively decompose ship-radiated noise signals \cite{ref29} and thus can be employed for more accurate DEMON analysis. Considering the non-uniform modulation characteristics of ship-radiated noise, we decompose the signal into multiple components using VMD and extract DEMON features from each component to form 2-D features for target recognition. The key parameters of VMD, namely the number of modes $K$ and the penalty factor $\alpha$, are determined based on prior knowledge and a small subset of samples. The detailed procedure is provided in \textbf{\ref{app1}}.

\subsubsection{3/2-D spectrum}
The 3/2-D spectrum of random process $x(t)$ is defined as the Fourier transform of the diagonal slice of its third-order cumulant $c_{3x}(\tau, \tau)$, yielding
\begin{equation}\label{eq3}
\begin{aligned}
C(\omega)&=\int_{-\infty}^{+\infty}{c_{3x}(\tau, \tau)e^{-j\omega\tau}}\,{\rm d}\tau\\ 
&=\int_{-\infty}^{+\infty}{\left[\int_{-\infty}^{+\infty}{x(t)x^2(t+\tau)\,{\rm d}t}\right]e^{-j\omega\tau}}\,{\rm d}\tau\\ 
&=X^{*}(\omega)\left[X(\omega)*X(\omega)\right]
\end{aligned}
\end{equation}
where $X(\omega)$ denotes the Fourier transform of $x(t)$, and $X^{*}(\omega)$ is the complex conjugate of $X(\omega)$. In this study, the calculation progress of 3/2-D spectrum is as follows:
\begin{enumerate}[(1)]
\item Divide the sample $x(n)$ into $L$ segments, each with a length of $N$, and remove the DC component.
\item Calculate the diagonal slice of the third-order cumulant in each segment:
\begin{equation}\label{eq4}
c^{(i)}(\tau)= \frac{1}{N}\sum_{n=n_1}^{n_2}x^{(i)}(n)x^{(i)}(n+\tau)x^{(i)}(n+\tau)
\end{equation}
where $i=1,2,...,L$, $n_1=\mathrm{max}(0, -\tau)$, and $n_2=\mathrm{min}(N-1, N-1-\tau)$
\item Compute the average value of $c^{(i)}(\tau)$:
\begin{equation}\label{eq5}
\hat{c}(\tau)=\frac{1}{L}\sum_{i=1}^{L}c^{(i)}(\tau)
\end{equation}
\item Calculate the discrete Fourier transform of $\hat{c}(\tau)$, and finally obtain the 3/2-D spectrum of $x(n)$.
\end{enumerate}

While suppressing Gaussian noise and eliminating non-phase-coupled harmonic terms, the 3/2-D spectrum inevitably leads to some information loss. As reported in \cite{ref11}, using only the bispectrum yields lower recognition performance than using only the amplitude spectrum. Therefore, we propose a feature fusion method that employs the 3/2-D DEMON spectrum to enhance key frequency information while retaining the remaining information. 
\begin{figure}[htbp]
  % \vspace{-0.1cm}
  \centering
  \includegraphics[width=0.8\textwidth]{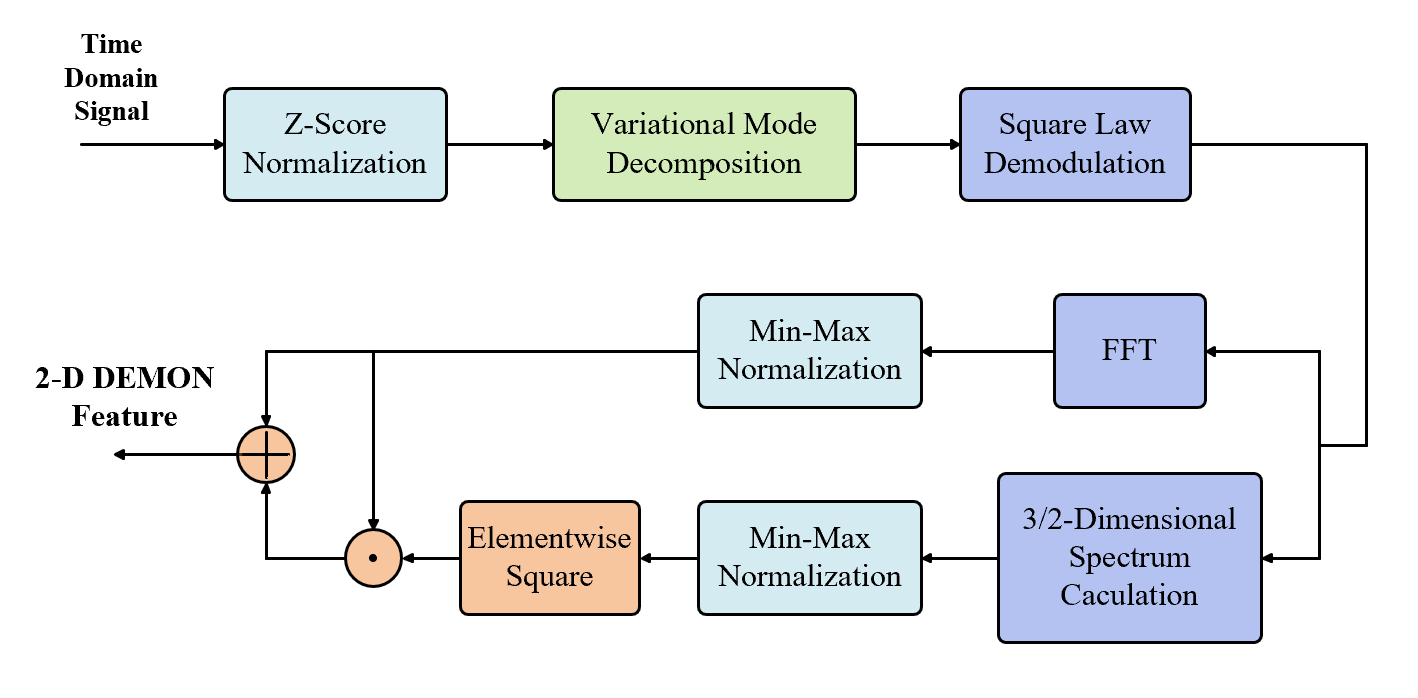} 
  \caption{2-D DEMON spectral feature extraction and fusion process}
  \label{fig1}
  % \vspace{-0.3cm}
\end{figure}

\begin{figure}[htbp]
  \vspace{-0.2cm}
  \centering
  \begin{subfigure}[b]{.32\linewidth}
    \centering
    \includegraphics[width=\linewidth]{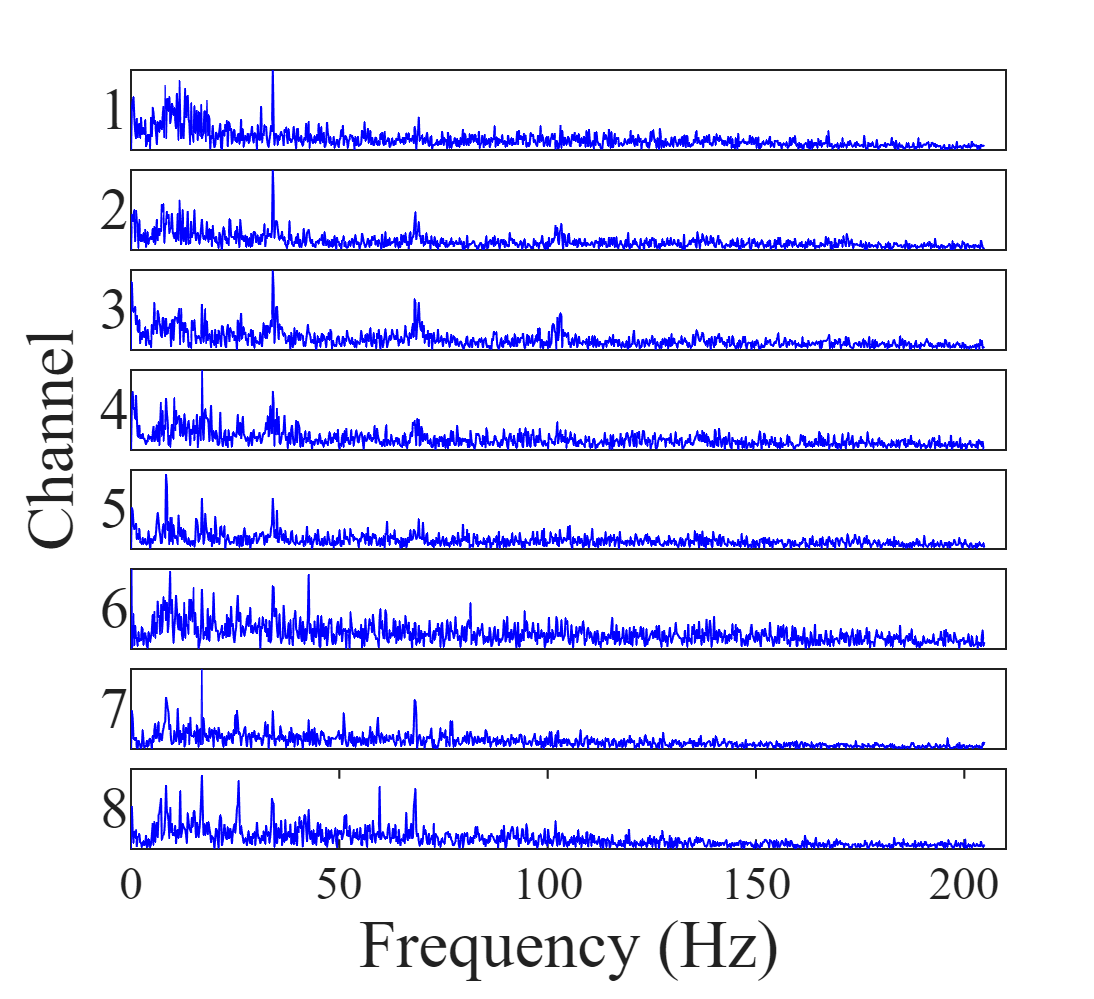}
    \caption{Amplitude spectrum}
    \label{subfig1}
  \end{subfigure}
  \begin{subfigure}[b]{.32\linewidth}
    \centering
    \includegraphics[width=\linewidth]{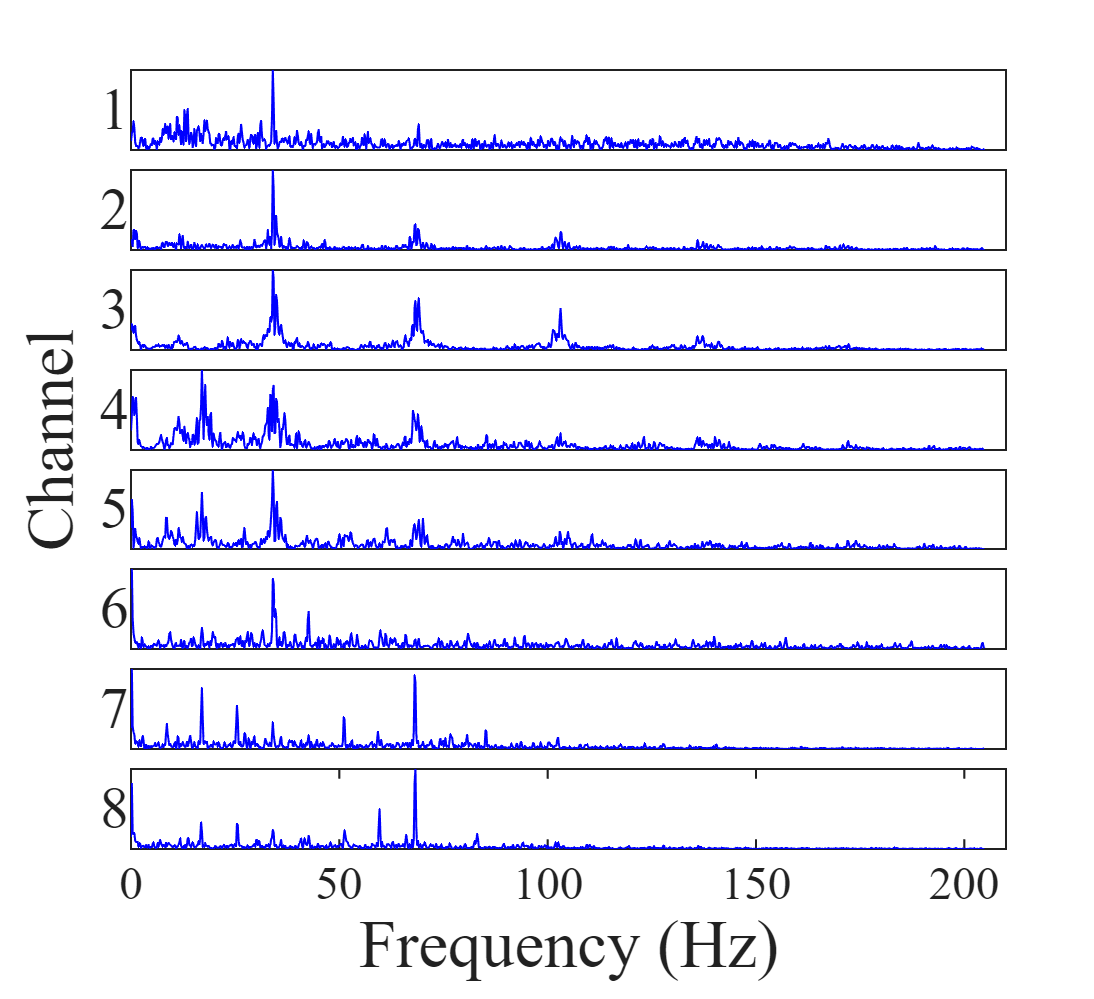}
    \caption{3/2-D spectrum}
    \label{subfig2}
  \end{subfigure}  
  \begin{subfigure}[b]{.32\linewidth}
    \centering
    \includegraphics[width=\linewidth]{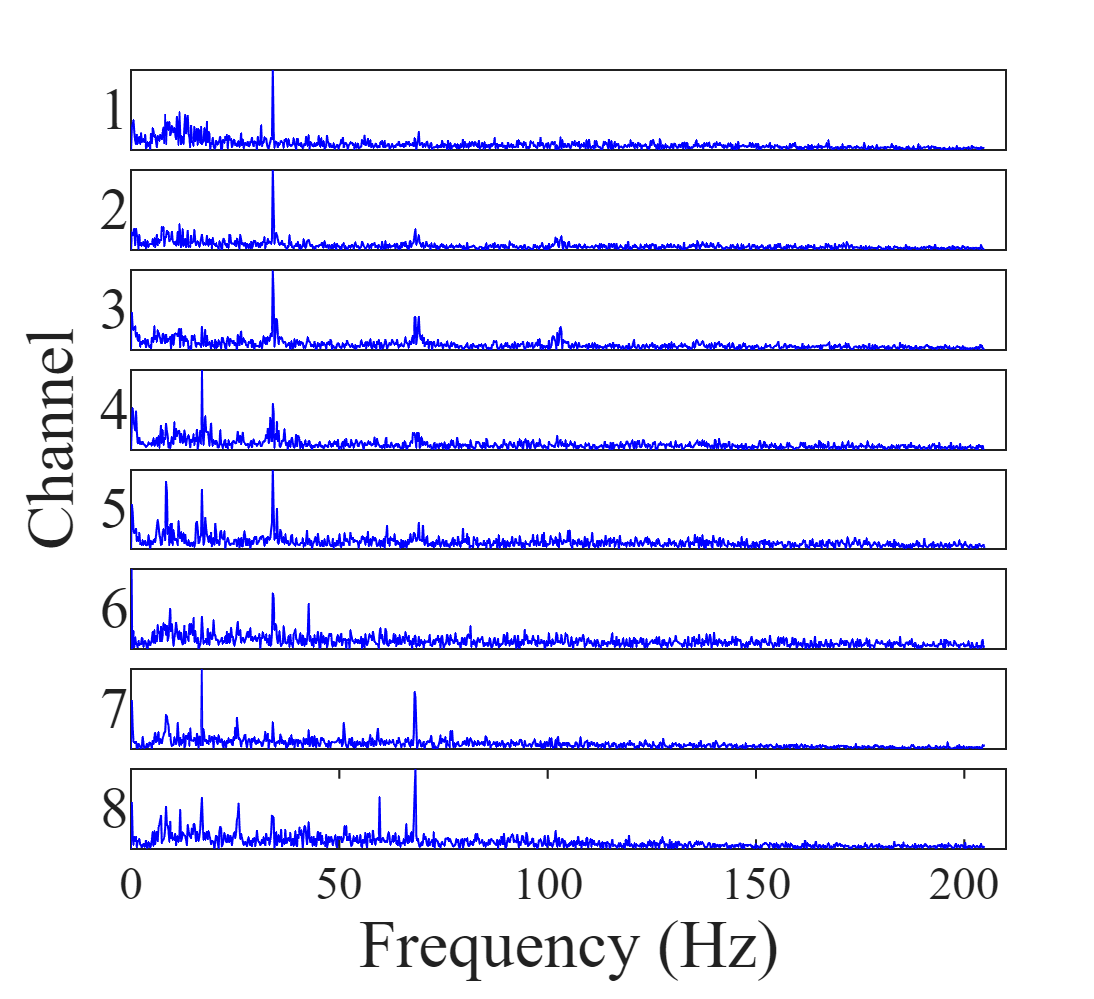}
    \caption{Fusion spectrum}
    \label{subfig3}
  \end{subfigure}
  \caption{2-D DEMON spectral feature}
  \label{fig2}
\end{figure}

Fig. \ref{fig1} illustrates the proposed feature extraction and fusion process. First, the normalized signal is decomposed into multiple IMF components using VMD. For each IMF component, square-law demodulation is applied to extract the envelope information, while both the amplitude spectrum and the 3/2-D spectrum are computed. After min-max normalization, the product of the squared 3/2-D spectrum and the amplitude spectrum is calculated and then added to the amplitude spectrum to achieve feature fusion. Finally, the 1-D envelope spectra obtained from all components are concatenated to form the final 2-D DEMON spectral feature. Fig. \ref{fig2} presents examples of the DEMON amplitude spectrum, DEMON 3/2-D spectrum, and the resulting 2-D fused DEMON spectral feature of ship-radiated noise. Specifically, Fig. \ref{subfig1} demonstrates the modulation inhomogeneity. Moreover, the suppression of noise and non-phase-coupled frequency components in the DEMON 3/2-D spectrum is clearly observed. Additionally, it can be observed that the fused spectrum highlights the critical line-spectrum frequency components while retaining certain detailed information from the amplitude spectrum.

\subsection{Network architecture}
\label{subsec2}
\begin{figure}[htbp]
  % \vspace{-0.1cm}
  \centering
  \includegraphics[width=\textwidth]{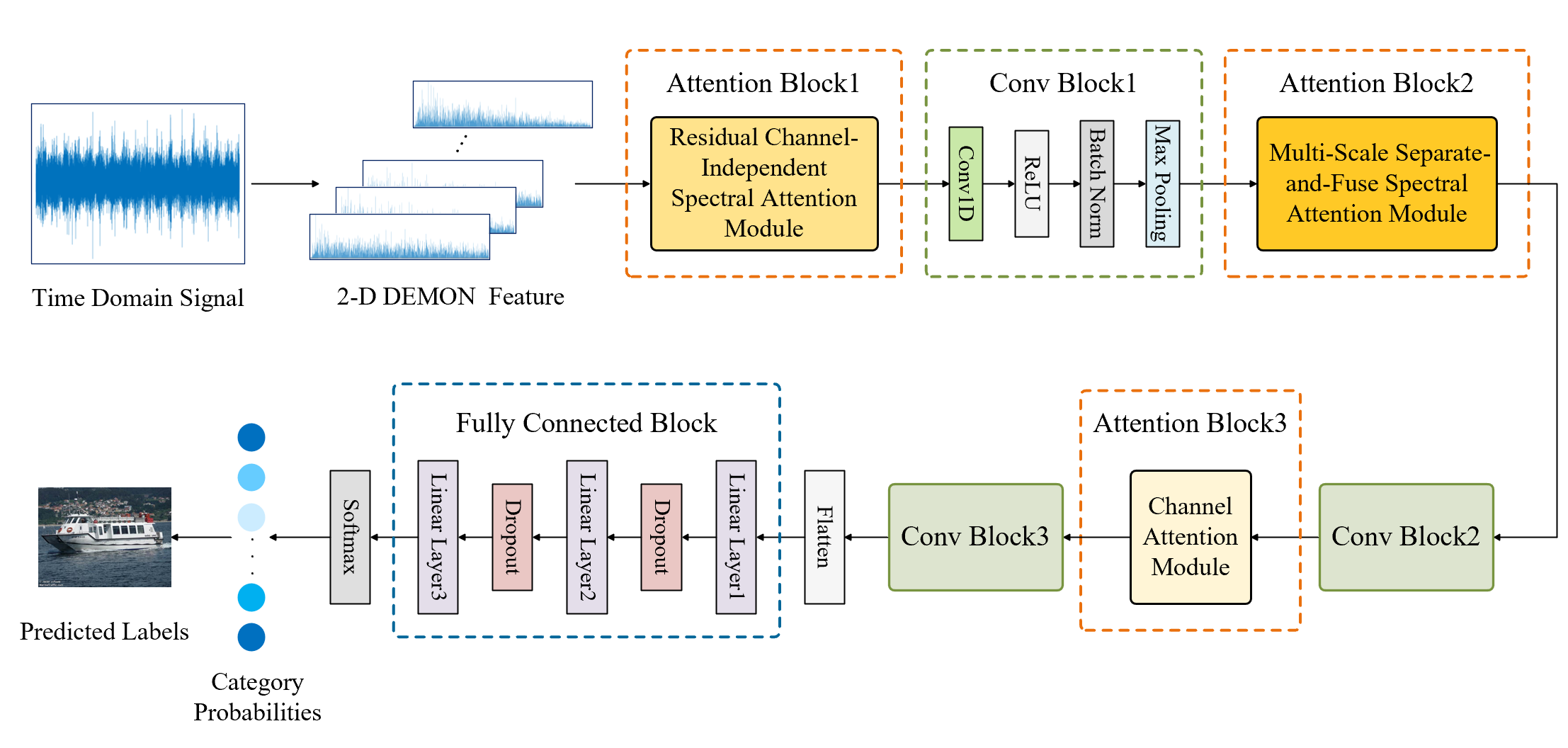} 
  \caption{The framework of the recognition model}
  \label{fig3}
  % \vspace{-0.3cm}
\end{figure}

In this study, the 1-D CNN is employed to construct the recognition model, in which three representative attention mechanisms are incorporated. The architecture of the proposed model is illustrated in Fig. \ref{fig3}.

The 2-D DEMON spectral features extracted in Section \ref{subsec1} serve as the input to the 1-D CNN. As shown in Fig. \ref{fig3}, the attention modules have diverse structures, which are detailed in Section \ref{subsec3}. All convolutional modules share the same  fundamental architecture, including a 1-D convolutional layer, an activation layer (ReLU), a batch normalization layer, and a pooling layer (max-pooling). Batch normalization is applied to prevent vanishing and exploding gradients, thereby accelerating convergence and improving stability. In the \textbf{Fully Connected Block}, the first two linear layers adopt the ReLU activation function, while the output of the final linear layer is fed into a softmax function to generate class probability predictions. Dropout layers are added after the first two fully connected layers as a regularization measure to reduce overfitting. 

\begin{table}[htbp]
  % \begin{center}
  \caption{The parameters of the recognition network}
  \centering
  \fontsize{8}{12}\selectfont
  % \footnotesize
  \scalebox{0.99}{
    \begin{tabular}{cccc}
      \hline
      \textbf{Module} & \textbf{Layer type} & \textbf{Output size} & \textbf{Specific parameter} \\
      \hline
      \multirow{2}{*}{Attention Block 1} & \multirow{2}{*}{R-CISAM} & \multirow{2}{*}{$8 \times 1024$} & $\mathrm{kernel \enspace size}=3,\enspace \mathrm{stride}=1$ \\
       & & & $\mathrm{dropout \enspace probability}=0.8$ \\
      \hline
      \multirow{4}{*}{\text{Conv Block 1}} & \text{Conv1D} & $32 \times 1024$ & $\mathrm{kernel \enspace size}=3 ,\enspace \mathrm{stride}=1$ \\
      & \text{ReLU} & $32 \times 1024$ & {} \\
      & \text{BatchNorm} & $32 \times 1024$ & {} \\
      & \text{MaxPooling} & $32 \times 512$ & $ \mathrm{kernel \enspace size}=2,\enspace \mathrm{stride}=2$ \\
      \hline
      \text{Attention Block 2} & \text{MS-SFSAM} & $32 \times 512$ & $ \mathrm{dilation \enspace rate}=[4, 8, 16],\enspace \mathrm{stride}=1$ \\
      \hline
      \multirow{4}{*}{\text{Conv Block 2}} & \text{Conv1D} & $64 \times 512$ & $\mathrm{kernel \enspace size}=3,\enspace \mathrm{stride}=1$ \\
      & \text{ReLU} & $64 \times 512$ & {} \\
      & \text{BatchNorm} & $ 64 \times 512$ & {} \\
      & \text{MaxPooling} & $ 64 \times 256$ & $\mathrm{kernel \enspace size}=2,\enspace \mathrm{stride}=2$ \\
      \hline
      \text{Attention Block 3} & \text{CAM} & $64 \times 256$ & $\mathrm{reduction \enspace ratio}=2/8$ \\
      \hline
      \multirow{4}{*}{\text{Conv Block 3}} & \text{Conv1D} & $ 64 \times 256$ & $\mathrm{kernel \enspace size}=3,\enspace \mathrm{stride}=1$ \\
      & \text{ReLU} & $ 64 \times 256 $ & {} \\
      & \text{BatchNorm} & $ 64 \times 256 $ & {} \\
      & \text{MaxPooling} & $ 64 \times 128 $ & $ \mathrm{kernel \enspace size}=2,\enspace \mathrm{stride}=2$ \\
      \hline
      \text{Flatten} & {} & $8192$ & {} \\
      \hline      
      \multirow{2}{*}{\text{FC1}} & \text{Linear} & $1024$ & {} \\
      & \text{Dropout} & $1024$ & $\mathrm{dropout \enspace probability}=0.5$ \\
      \hline
      \multirow{2}{*}{\text{FC2}} & \text{Linear} & $128$ & {} \\
      & \text{Dropout} & $128$ & $\mathrm{dropout \enspace probability}=0.2$ \\
      \hline
      \text{FC3} & \text{Linear} & $12$ & {} \\
      \hline
    \end{tabular}
  % \end{center}
  }  
  \label{table2}
\end{table}

The main hyperparameters of each layer in the recognition network are listed in Table \ref{table2}, with those of the attention blocks determined via 5-fold cross-validation. The convolutional and pooling layers share the same architecture across all convolutional modules. Specifically, each convolutional layer uses a kernel size of $3$, a stride of $1$, and padding of $1$, while each max-pooling layer uses a pooling kernel size of $2$ and a stride of $2$. The three fully connected layers have output dimensions of $1024$, $128$, and $12$, respectively, and the two dropout layers have dropout rates of $0.5$ and $0.2$, respectively.

\subsection{Multi-Stage Multi-Type Attention Mechanism}
\label{subsec3}
In this work, we introduce a Multi-Stage Multi-Type Attention Mechanism tailored to the characteristics of 2-D DEMON spectral features and the classification network. The MMATT consists of three attention modules, which are placed before the three convolutional blocks, respectively. Moreover, it employs attention mechanisms in both frequency and channel domains of the feature maps. Specifically, \textbf{Attention Block 1} adopts the proposed R-CISAM, \textbf{Attention Block 2} employs the designed MS-SFSAM, and \textbf{Attention Block 3} utilizes a channel attention mechanism.

\subsubsection{Residual Channel-Independent Spectral Attention Mechanism}
\label{subsubsec1}
The frequencies and intensities of modulating signals vary across different sub-band components of ship-radiated noise. Therefore, we apply distinct spectral attention weights to the DEMON spectrum of each sub-band component and avoid interference among components caused by pooling in the conventional SAM scheme. Furthermore, a residual connection is adopted to retain original feature information, while dropout is applied to the attention weights during training to mitigate overfitting. Based on the above, we propose R-CISAM and apply it in \textbf{Attention Block 1} to weight the input 2-D DEMON spectral features.
\begin{figure}[htbp]
  \vspace{-0.1cm}
  \centering
  \includegraphics[width=0.9\textwidth]{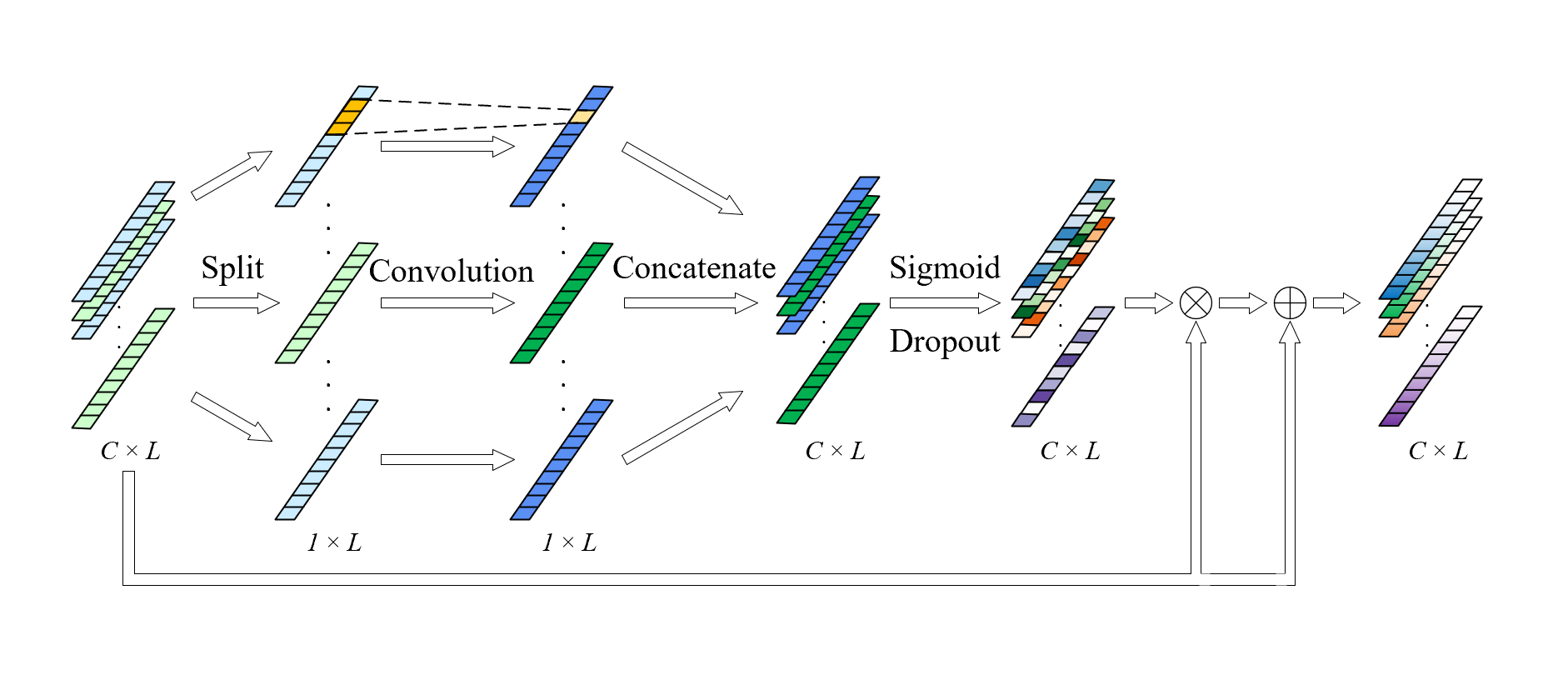}
  \caption{Residual Channel-Independent Spectral Attention Mechanism}
  \label{fig4}
  \vspace{-0.1cm}
\end{figure}
The architecture of the proposed R-CISAM module is shown in Fig. \ref{fig4}. First, depthwise convolution is applied to independently capture information of each channel. Next, a sigmoid function is used to generate frequency attention weights within the range $[0,1]$. Finally, the element-wise product of the attention weight and the input data is computed and then added back to the input. In contrast to the traditional SAM, R-CISAM is better suited for 2-D DEMON spectral features. By dynamically enhancing the envelope spectral features of ship-radiated noise, R-CISAM improves both recognition accuracy and robustness.

\subsubsection{Multi-Scale Separate-and-Fuse Attention Mechanism}
\label{subsubsec2}
R-CISAM involves no inter-channel information fusion and computes attention weights with a small receptive field, thereby focusing on the details of the original spectrum. In contrast, middle-layer features require global contextual relationships to supplement larger-scale information. Furthermore, different target types exhibit distinct scale characteristics. To address these requirements, we propose MS-SFSAM.

\begin{figure}[htbp]
  \centering
  \includegraphics[width=0.9\textwidth]{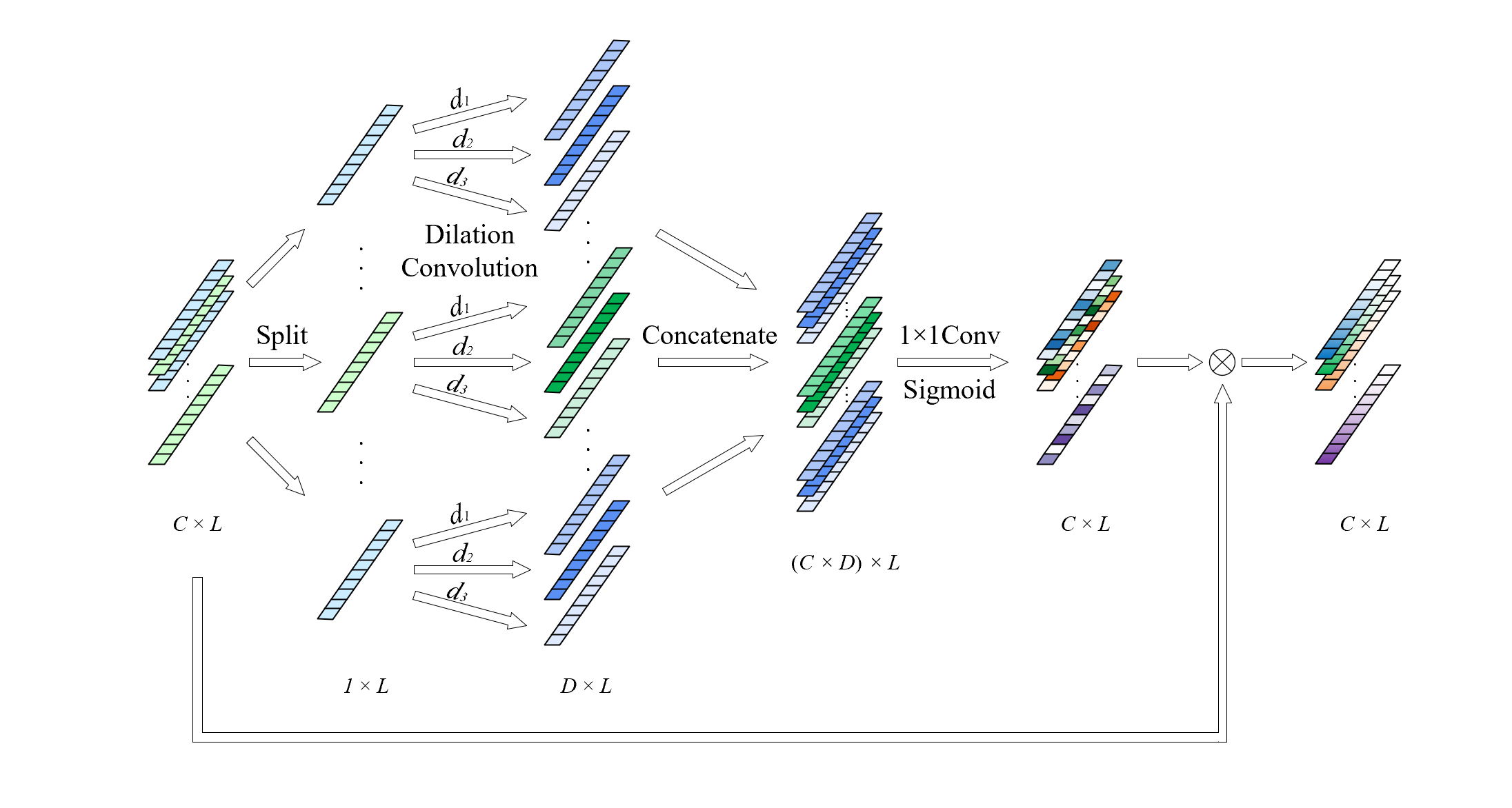} 
  \caption{Multi-Scale Separate-and-Fuse Spectral Attention Mechanism}
  \label{fig5}
\end{figure}

The multi-scale spectral attention weights are first computed per channel and then fused across channels. Fig. \ref{fig5} shows the architecture of the proposed MS-SFSAM. The attention weights are generated using dilated depthwise separable convolution. By applying R-CISAM and MS-SFSAM at different depths of the model, the multi-scale characteristics of the data are fully utilized, achieving hierarchical feature enhancement.

\subsubsection{Channel attention mechanism}
\label{subsubsec3}
\begin{figure}[htbp]
  \centering
  \includegraphics[width=0.75\textwidth]{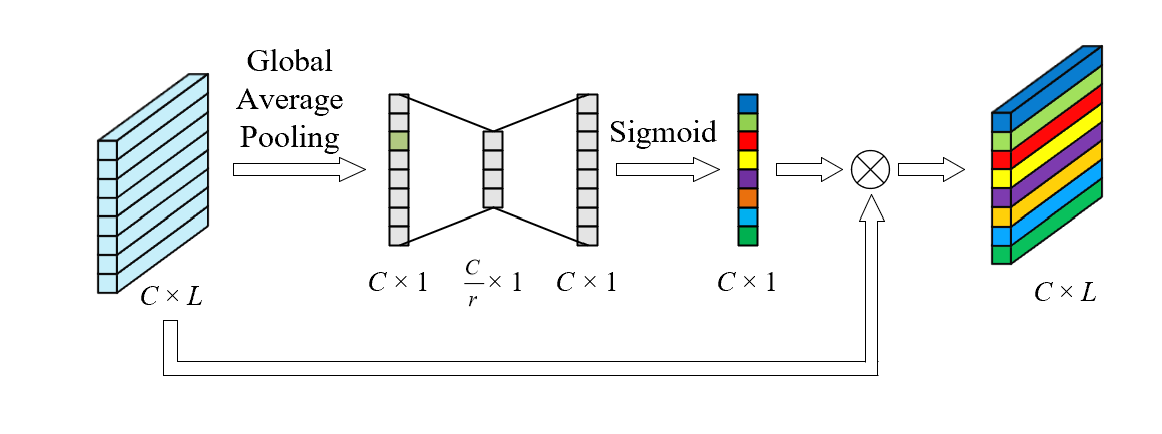} 
  \caption{Channel attention mechanism}
  % \vspace{-0.3cm}
\label{fig6}
\end{figure}

In our ship-radiated noise recognition model, the SE block \cite{ref30} is adopted as the channel attention module, as illustrated in Fig. \ref{fig6}. First, global average pooling is applied to aggregate global information for each channel. Then, two fully connected layers with output dimensions of $c/r$ and $c$ are used for activation, where $c$ denotes the number of channels and $r\in[1,c]$ represents the integer compression ratio. The two fully connected layers employ ReLU and sigmoid activation functions, respectively. Finally, element-wise multiplication is performed between the channel attention weights and the input data to enhance important channel information and suppress irrelevant information.

\subsection{Adjustable Class-Balanced Focal Loss}
\label{subsec4}
\subsubsection{Adjustable focal loss}
\label{subsubsec4}
The focal loss function proposed in \cite{ref25} can be given as
\begin{equation}\label{eq6}
\mathrm{FL} \left(p\right) = -\left(1 - p\right)^{\gamma} \mathrm{log} \left(p\right)
\end{equation}
where $p$ is the estimated probability of true class, $\gamma \geq 0$ denotes the tunable focusing parameter. For hard misclassified examples where $p$ is small, the focal loss approximates the cross-entropy loss. For easy well-classified examples with $p \rightarrow 1$, the focal loss becomes significantly smaller than the cross-entropy loss. By incorporating a dynamic modulating factor $\left(1-p\right)^{\gamma}$ into the cross-entropy loss, the focal loss assigns higher weights to examples with lower predicted probabilities, thereby focusing training on difficult examples. To generalize the focal loss to tasks with varying degrees of data imbalance, we propose an improved focal loss function, yielding
\begin{equation}\label{eq7}
\mathrm{AFL}\left(p\right) = -\left(1-\beta p\right)^{\gamma}\mathrm{log} \left(p\right)
\end{equation}
where $\beta \in \left[0, 1 \right]$ is the adjustable parameter. By adjusting the values of $\beta$, the loss function varies between cross-entropy loss and the original focal loss function. 

\subsubsection{Adjustable class-balanced loss function}
\label{subsubsec5}
To adapt class weights to the data distribution of specific tasks, we propose an adjustable class-balanced loss function by introducing a tunable parameter into the static class weights, yielding
\begin{equation}\label{eq8}
\mathrm{ACBL} \left(p_{y}\right) = \frac{M}{\sum_{i = 1}^{M} \frac{1}{m_{i}}} \left(\frac{1}{m_{y}}\right)^{q} \mathrm{Loss} \left(p_{y}\right)
\end{equation}
where $y$ denotes the true class of sample, $p_y$ is predicted probability, $m_i$ represents the number of samples belonging to class $i$, while $M$ is the number of categories, and $\mathrm{Loss}(p_y)$ represents the loss function without class balancing measurement. Furthermore, $q \in \left[0,1\right]$ is a class-balanced tunable parameter, $q = 0$ corresponds to no class-balanced weighting, and $q = 1$ implies that the weights are inversely proportional to class frequencies. Combining Eq. \eqref{eq7} and Eq. \eqref{eq8}, the proposed Adjustable Class-Balanced Focal Loss function can be expressed as
\begin{equation}\label{eq9}
\mathrm{ACBFL} \left(p_{y}\right) = - \frac{M}{\sum_{i = 1}^{M} \frac{1}{m_{i}}} \left(\frac{1}{m_{y}}\right)^{q} \left(1 - \beta p_{y}\right)^{\gamma} \mathrm{log} \left(p_{y}\right)
\end{equation}
where $q$, $\beta$, and $\gamma$ are adjustable parameters determined by the specific task and dataset.

\section{Experiment setup}
\label{sec3}
\subsection{Dataset}
\label{subsec5}
In this study, experiments are conducted on the open-source ShipsEar dataset\cite{ref28}. The audio data were collected in various areas along the Atlantic coast of northwest Spain during autumn $2012$ and summer $2013$, comprising $11$ types of ship-radiated noise as well as natural environment noise. The dataset contains $90$ recordings with durations ranging from $15$ s to $10$ min, and a sampling rate of $52734$ Hz. The audio data are first downsampled to one-third of the original sampling rate and then split into non-overlapping $5$-second fragments. Due to the small dataset size, we adopt 5-fold cross-validation to eliminate the influence of data distribution on the results. Furthermore, to mitigate the effect of randomness, repeated experiments are conducted using 5 random seeds within each fold. Table \ref{table3} details the number of training and test samples for each class.

\begin{table}[htbp]
  \centering
  \caption{The number of training and test samples for each category}
  % \begin{center}
    \label{table3}
    \fontsize{8}{12}\selectfont
    % \footnotesize
    \scalebox{0.99}{
    % \resizebox{0.8\textwidth}{!}{
    \begin{tabular}{cccccc}
      \hline
      \textbf{Class} & \textbf{Training sample} & \textbf{Test sample} & \textbf{Class} & \textbf{Training sample} &\textbf{Test sample} \\
      \hline
      \text{Dredger} & $41$ & $11$ & \text{Passenger} & $666$ & $166$ \\
      \text{Fish boat} & $81$ & $20$ & \text{Pilot} & $20$ & $6$ \\
      \text{Motorboat} & $153$ & $38$ & \text{Ro-Ro} & $240$ & $60$ \\
      \text{Mussel} & $115$ & $29$ & \text{Sailboat} & $64$ & $15$ \\
      \text{Natural noise} & $179$ & $45$ & \text{Trawler} & $25$ & $7$ \\
      \text{Ocean liner} & $149$ & $37$ & \text{Tugboat} & $32$ & $8$ \\
      \hline
    \end{tabular}
    % }
    }
  % \end{center}
\end{table}

\subsection{Implementation details}
\label{subsec6}
Experiments are conducted on the ShipsEar dataset to evaluate the proposed approaches and recognition model from three perspectives: feature extraction, attention mechanism, and loss function. Ablation experiments are also performed to assess the contribution of each attention block. Additionally, we analyze the impact of the parameters $q$ and $\beta$ in ACBFL on classification performance. All models are trained for $100$ epochs with a batch size of $32$ using the Adam optimizer. The initial learning rate is $10^{-3}$, and it decays by a factor of $0.6$ every $10$ epochs.

\subsection{Evaluation metrics}
\label{subsec7}
Due to class imbalance, we adopt three metrics to evaluate model performance comprehensively: overall accuracy (OA), F1-score, and average accuracy (AA). The formulas for these indicators are as follows:

\begin{equation}\label{eq10}
\mathrm{Overall \textrm{ } Accuracy} = \frac{\sum_{i = 1}^{M} \mathrm{TP}_{i}}{\sum_{i = 1}^{M} \left(\mathrm{TP}_{i} + \mathrm{FN}_{i}\right)}
\end{equation}

\begin{equation}\label{eq11}
\mathrm{Precision}_{i} = \frac{\mathrm{TP}_{i}}{\mathrm{TP}_{i} + \mathrm{FP}_{i}}
\end{equation}

\begin{equation}\label{eq12}
\mathrm{Recall}_{i} = \frac{\mathrm{TP}_{i}}{\mathrm{TP}_{i} + \mathrm{FN}_{i}}
\end{equation}

\begin{equation}\label{eq13}
\mathrm{F1\mbox{-}score}_{i} = 2 \cdot \frac{\mathrm{Precision}_{i} \cdot \mathrm{Recall}_{i}}{\mathrm{Precision}_{i} + \mathrm{Recall}_{i}}
\end{equation}

\begin{equation}\label{eq14}
\mathrm{F1\mbox{-}score} = \frac{1}{M} \sum_{i = 1}^{M} \mathrm{F1\mbox{-}score}_{i}
\end{equation}

\begin{equation}\label{eq15}
\mathrm{Average \textrm{ } Accuracy} = \frac{1}{M} \sum_{i = 1}^{M} \mathrm{Recall}_{i}
\end{equation}
where $M$ is the number of categories, $\mathrm{TP}_i$, $\mathrm{FP}_i$, and $\mathrm{FN}_i$ represent true positive, false positive, and false negative of class $i$, respectively. Moreover, $\mathrm{Precision}_i$, $\mathrm{Recall}_i$, and $\mathrm{F1\mbox{-}score}_i$ represent the precision, recall, and F1-score of class $i$.

\section{Experiments results}
\label{sec4}
The mean and standard deviation across the 5 folds are reported as the final results. All experiments are performed using Python $3.13.9$ and Pytorch $2.9.1$ on a computer equipped with a Core i$7$-$14650$HX CPU and an NVIDIA GeForce RTX $5060$ GPU.

\subsection{Comparison of feature extraction methods}
\label{subsec8}
In this subsection, we compare the classification performance of models using different input features. The basic backbone network and cross-entropy loss are adopted. Table \ref{table4} reports the results, where Trad-CNN utilizes the original 1-D DEMON feature; Filter-CNN uses the 2-D DEMON feature whose sub-band components are obtained via band-pass filtering; VMD-CNN and VMD-3/2D-CNN invoke the VMD-based 2-D DEMON spectrum and the 3/2-D spectrum, respectively; and VMD-Fusion-CNN exploits our VMD-based fusion 2-D DEMON feature. Among all three evaluation metrics, VMD-Fusion-CNN achieves the highest values. Specifically, its OA, F1-score, and AA exceed those of Filter-CNN by $4.43\%$, $4.2\%$, $5.6\%$ and those of Trad-CNN by $16.29\%$, $21.11\%$, $22.2\%$, respectively.

\begin{table}[htbp]
  % \footnotesize
  \begin{center}
    \caption{Results of different feature extraction methods}
    \label{table4}
    \fontsize{9}{12}\selectfont
    \scalebox{0.99}{
    \begin{tabular}{cccc}
      \hline
      \textbf{Method} & \textbf{Overall Accuracy (\%)} & \textbf{F1-score (\%)} & \textbf{Average Accuracy (\%)} \\
      \hline
      \text{Trad-CNN} & $73.21 \pm 1.45$ & $63.79 \pm 1.82$ & $60.38 \pm 1.50$ \\
      \text{Filter-CNN} & $85.07 \pm 1.25$ & $80.70 \pm 2.40$ & $76.98 \pm 2.41$ \\
      \text{VMD-CNN} & $88.71 \pm 0.98$ & $84.43 \pm 0.58$ & $81.62 \pm 1.52$ \\
      \text{VMD-3/2D-CNN} & $81.65 \pm 0.73$ & $74.30 \pm 2.52$ & $70.70 \pm 2.38$ \\
      \textbf{VMD-Fusion-CNN} & $\mathbf{89.50 \pm 0.72}$ & $\mathbf{84.90 \pm 1.08}$ & $\mathbf{82.58 \pm 1.75}$ \\
      \hline
    \end{tabular}
    }
  \end{center}
  \vspace{-0.3cm}
\end{table}

\begin{figure}[htbp]
  % \vspace{-0.1cm}
  \centering
  \begin{subfigure}[b]{.28\linewidth}
    \centering
    \includegraphics[width=\linewidth]{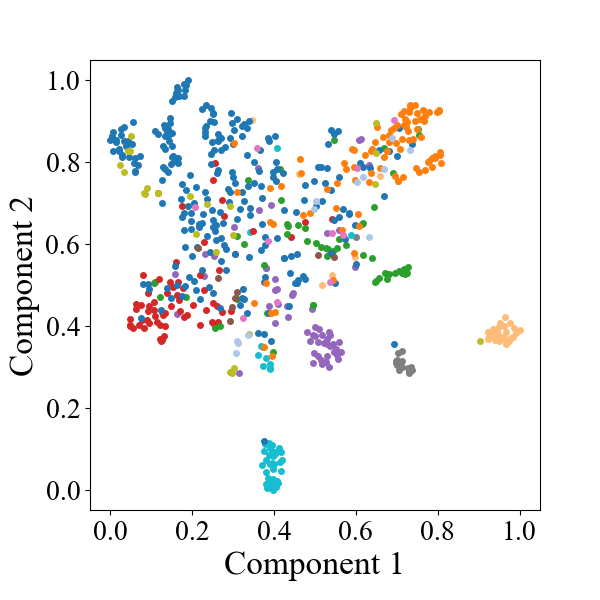}
    \caption{Trad-CNN}
    \label{subfig4}
  \end{subfigure}
  \begin{subfigure}[b]{.28\linewidth}
    \centering
    \includegraphics[width=\linewidth]{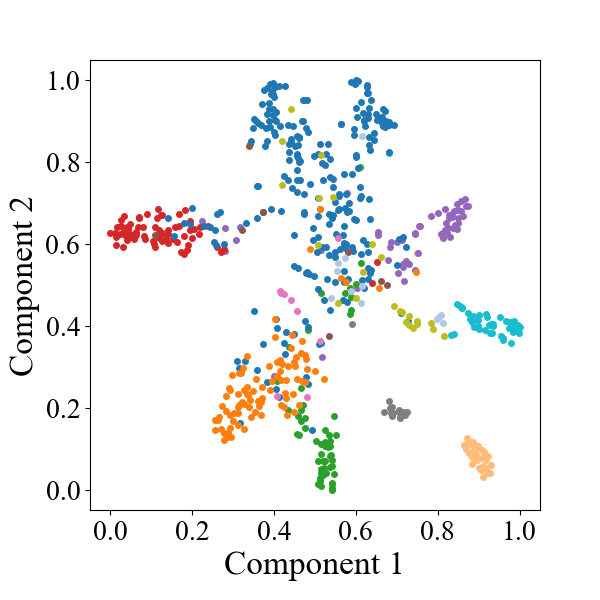}
    \caption{Filter-CNN}
    \label{subfig5}
  \end{subfigure}  
  \begin{subfigure}[b]{.28\linewidth}
    \centering
    \includegraphics[width=\linewidth]{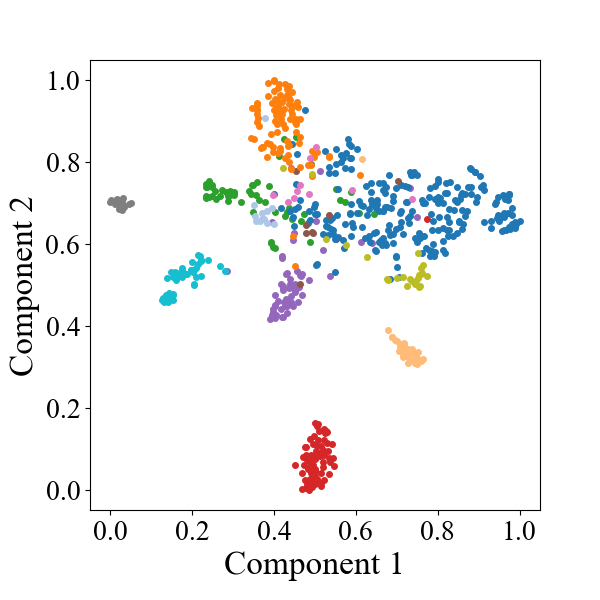}
    \caption{VMD-Fusion-CNN}
    \label{subfig6}
  \end{subfigure}
  \begin{subfigure}[b]{.12\linewidth}
    \centering
    \includegraphics[width=\linewidth]{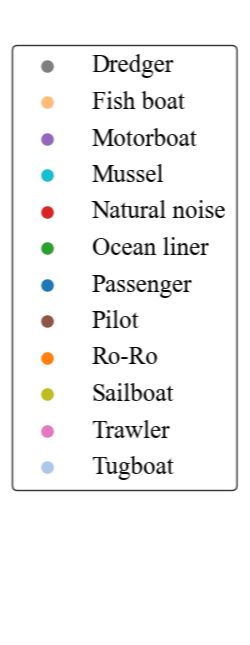}
    % \caption{Legend}
    % \label{subfig6}
  \end{subfigure}
  \caption{The distribution of deep features of three models with different input features}
  \label{fig7}
\end{figure}

In addition, we extract deep features from the penultimate fully connected layer and visualize their distribution using t-distributed stochastic neighbor embedding (t-SNE) algorithm \cite{ref31}, as shown in Fig. \ref{fig7}. The deep features obtained by VMD-Fusion-CNN exhibit superior inter-class separability and intra-class compactness compared to other counterparts. These results indicate that our enhanced 2-D DEMON feature extraction and fusion method significantly outperforms the original DEMON solution and also achieves better performance than existing alternatives. We additionally compare the model using time-frequency spectrogram, the most commonly used input features in recognition methods, with further details provided in \textbf{\ref{app2}}.

\subsection{Recognition model evaluation}
\label{subsec9}
\begin{figure}[htbp] 
  % \vspace{-0.2cm}
  \centering
  \begin{subfigure}[b]{.48\linewidth}
    \centering
    \includegraphics[width=\linewidth]{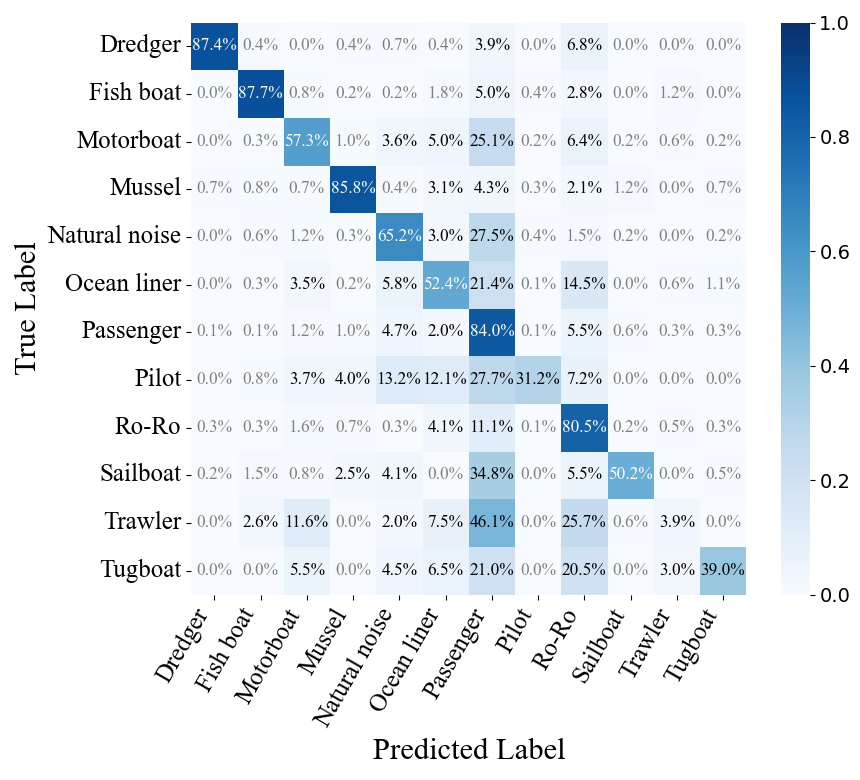}
    \caption{Baseline(Trad-CNN)}
    \label{subfig7}
  \end{subfigure}
  \begin{subfigure}[b]{.48\linewidth}
    \centering
    \includegraphics[width=\linewidth]{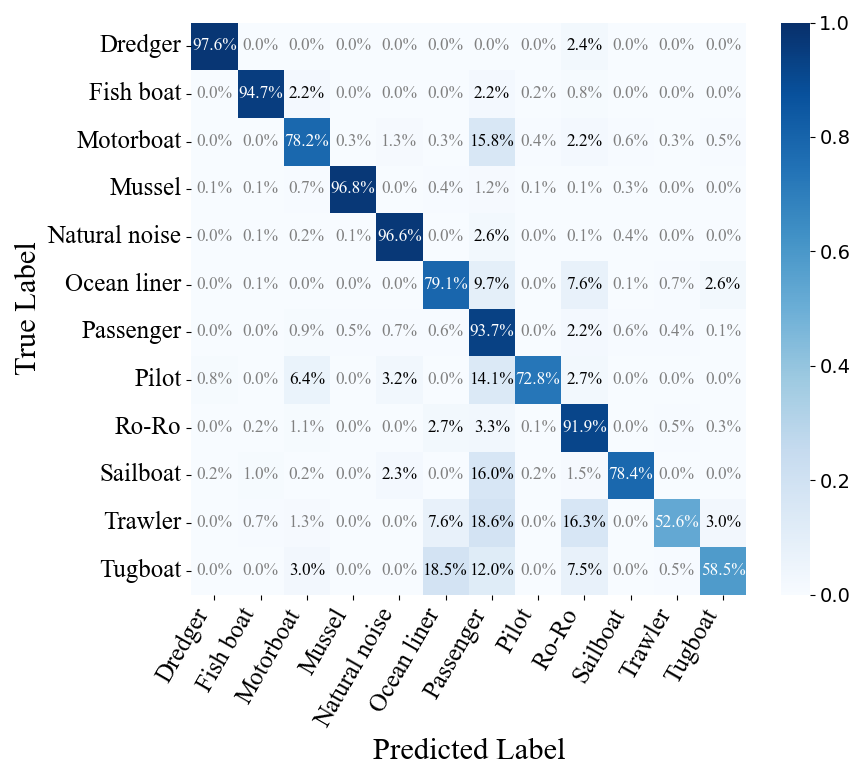}
    \caption{VMD-Fusion-CNN}
    \label{subfig8}
  \end{subfigure}  
  \begin{subfigure}[b]{.48\linewidth}
    \centering
    \includegraphics[width=\linewidth]{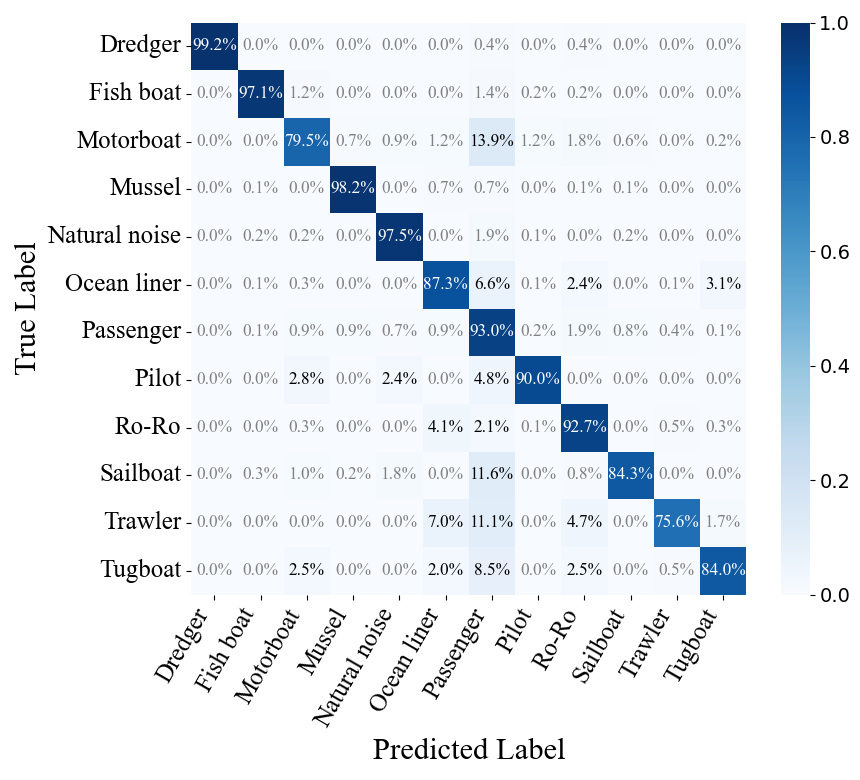}
    \caption{VMD-Fusion-CNN-MMATT-ImFL}
    \label{subfig9}
  \end{subfigure}
  \caption{Confusion matrices of three recognition models}
  % \vspace{-0.2cm}
  \label{fig8}
\end{figure}

To evaluate model performance on long-tailed ship-radiated noise data, we conduct comparative experiments. The Trad-CNN model from Section \ref{subsec8} is used as the baseline. Our VMD-Fusion-CNN-MMATT-ImFL integrates three components: VMD-Fusion (VMD-based fused 2-D DEMON input), MMATT (Multi-Stage Multi-Type Attention Mechanism), and ImFL (improved focal loss, i.e., ACBFL). Fig. \ref{fig8} presents confusion matrices for Trad-CNN, VMD-Fusion-CNN, and our full model. Our model achieves the best recognition performance.

Table \ref{table5} reports the results of the comparative experiments. Our proposed model significantly outperforms the baseline, with improvements of $18.47\%$, $26.25\%$, and $29.49\%$ in OA, F1-score, and AA, respectively, verifying its effectiveness. Using the same input and network architecture as our proposed model, but with cross-entropy loss, VMD-Fusion-CNN-MMATT-CE improves over the VMD-Fusion-CNN model by $2.05\%$, $3.96\%$, and $4.7\%$ in the three metrics, demonstrating the benefit of MMATT. Furthermore, VMD-Fusion-CNN-MMATT-ImFL achieves substantially higher F1-score and AA than VMD-Fusion-CNN-MMATT-CE, indicating that ACBFL effectively mitigates class and hard-easy sample imbalances, making it more suitable for long-tailed data. To evaluate the individual contributions of R-CISAM and MS-SFSAM, we replace them with SAM, resulting in the VMD-Fusion-CNN-MMATTC1-ImFL and VMD-Fusion-CNN-MMATTC2-ImFL models. Our full model outperforms these two variants by $2.05\%$, $3.6\%$, $5.2\%$ and $0.42\%$, $0.98\%$, $1.65\%$ in OA, F1-score, and AA, respectively, confirming that both proposed attention mechanisms are more effective than the traditional SAM.

\begin{table}[htbp]
  % \footnotesize
  \begin{center}
    \caption{Results of the comparative experiment}
    \label{table5}
    \fontsize{8}{12}\selectfont
    \scalebox{0.99}{
    \begin{tabular}{cccc}
      \hline
      \textbf{Method} & \textbf{Overall Accuracy (\%)} & \textbf{F1-score (\%)} & \textbf{Average Accuracy (\%)} \\
      \hline
      \text{Baseline} & $73.21 \pm 1.45$ & $63.79 \pm 1.82$ & $60.38 \pm 1.50$ \\
      \text{VMD-Fusion-CNN} & $89.50 \pm 0.72$ & $84.90 \pm 1.08$ & $82.58 \pm 1.75$ \\
      \text{VMD-Fusion-CNN-MMATT-CE} & $91.55 \pm 0.99$ & $88.86 \pm 0.81$ & $87.28 \pm 0.96$ \\
      \textbf{VMD-Fusion-CNN-MMATT-ImFL} & $\mathbf{91.68 \pm 1.00}$ & $\mathbf{90.04 \pm 0.76}$ & $\mathbf{89.87 \pm 0.63}$ \\
      \text{VMD-Fusion-CNN-MMATTC1-ImFL} & $89.63 \pm 0.88$ & $86.44 \pm 0.82$ & $84.67 \pm 0.87$ \\
      \text{VMD-Fusion-CNN-MMATTC2-ImFL} & $91.26 \pm 0.93$ & $89.06 \pm 0.78$ & $88.22 \pm 0.82$ \\
      \hline
    \end{tabular}
    }
  \end{center}
  \vspace{-0.5cm}
\end{table}

\begin{figure}[htbp]
  % \vspace{-0.3cm}
  \centering
  \includegraphics[width=0.6\textwidth]{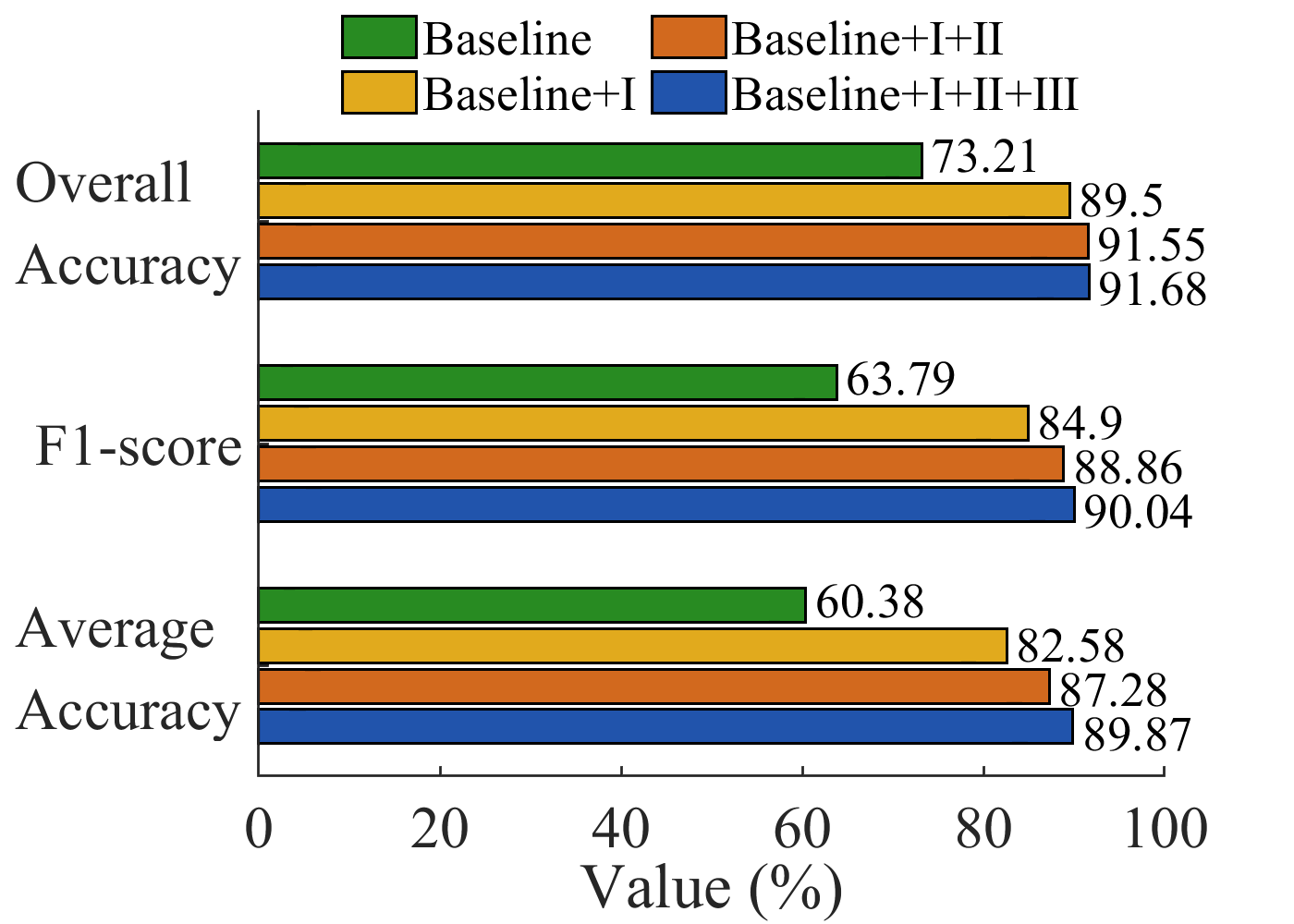} 
  \caption{The results of comparative experiments}
  \label{fig9}
\end{figure}

Fig. \ref{fig9} illustrates the recognition performance of models incorporating our proposed improvements, demonstrating the effectiveness of each component: \uppercase\expandafter{\romannumeral1} (VMD-based fusion 2-D DEMON feature), \uppercase\expandafter{\romannumeral2} (MMATT), and \uppercase\expandafter{\romannumeral3} (ACBFL).

The MS-SFSAM employs multi-scale dilation rates of 4, 8, and 16. We compare it with models using a single dilation rate, and Table \ref{table6} presents the experimental results. Specifically, the models VMD-Fusion-CNN-MMATTD1-ImFL, VMD-Fusion-CNN-MMATTD2-ImFL, and VMD-Fusion-CNN-MMATTD3-ImFL use dilation rates of 4, 8, and 16, respectively. The results show that the multi-scale approach outperforms any single-dilation-rate model, indicating that features at different scales are complementary and jointly contribute to improved performance.

\begin{figure}[htb]
  \vspace{-0.3cm}
  \centering
  \includegraphics[width=0.6\textwidth]{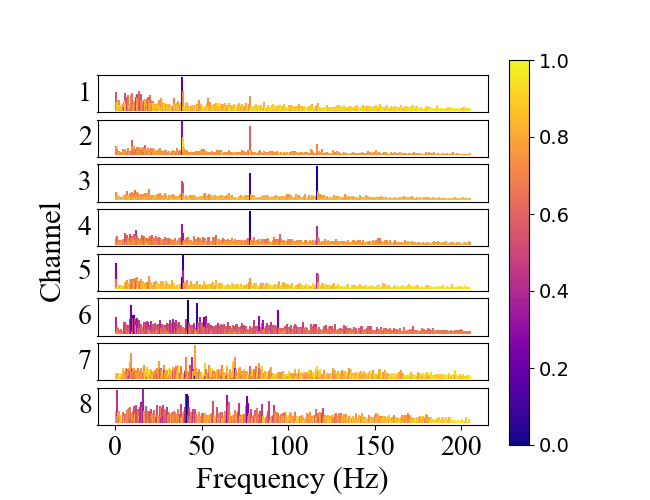} 
  \caption{Heat map of attention weights for R-CISAM}
  \label{fig10}
\end{figure}

\begin{table}[htbp]
  % \footnotesize
  % \vspace{-0.1cm}
  \begin{center}
    \caption{Experimental results with different dilation rates}
    \label{table6}
    \fontsize{8}{12}\selectfont
    \scalebox{0.99}{
    \begin{tabular}{cccc}
      \hline
      \textbf{Method} & \textbf{Overall Accuracy (\%)} & \textbf{F1-score (\%)} & \textbf{Average Accuracy (\%)} \\
      \hline
      \textbf{VMD-Fusion-CNN-MMATT-ImFL} & $\mathbf{91.68 \pm 1.00}$ & $\mathbf{90.04 \pm 0.76}$ & $\mathbf{89.87 \pm 0.63}$ \\
      \text{VMD-Fusion-CNN-MMATTD1-ImFL} & $89.93 \pm 0.94$ & $86.68 \pm 1.14$ & $84.58 \pm 1.44$ \\
      \text{VMD-Fusion-CNN-MMATTD2-ImFL} & $91.32 \pm 0.84$ & $89.02 \pm 0.46$ & $88.17 \pm 1.05$ \\
      \text{VMD-Fusion-CNN-MMATTD3-ImFL} & $91.16 \pm 1.22$ & $88.81 \pm 1.04$ & $89.03 \pm 0.70$ \\
      \hline
    \end{tabular}
    }
  \end{center}
  \vspace{-0.5cm}
\end{table}
To directly observe the effect of R-CISAM, we visualize its attention weights using a single input sample. As shown in Fig. \ref{fig10}, R-CISAM effectively discriminates among different frequencies. Furthermore, the varying distributions of attention weights across different channels validate the rationality of R-CISAM.

\subsection{Ablation experiments}
\label{subsec10}

\begin{figure}[htbp]
  \vspace{-0.3cm}
  \centering
  \includegraphics[width=0.7\textwidth]{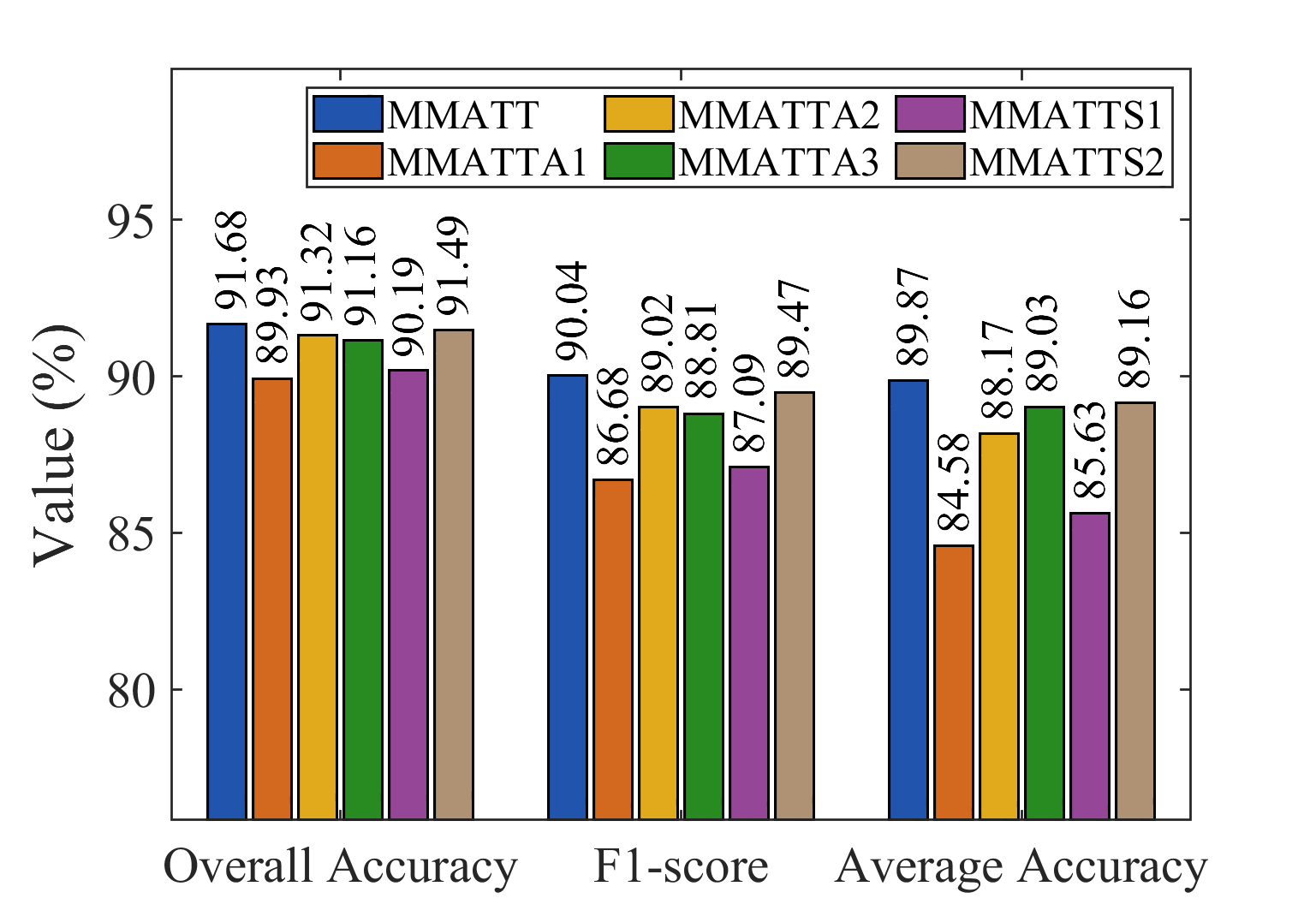}
  \caption{Recognition results of the ablation experiments}
  \label{fig11}
\end{figure}

To evaluate the contribution of each attention module in MMATT, we conduct ablation experiments. Specifically, three variants are derived from the full model VMD-Fusion-CNN-MMATT-ImFL by removing \textbf{Attention Block 1, Block 2, and Block 3}, respectively, resulting in VMD-Fusion-CNN-MMATTA1-ImFL, VMD-Fusion-CNN-MMATTA2-ImFL, and VMD-Fusion-CNN-MMATTA3-ImFL. Table \ref{table7} lists the results of the ablation experiments, and Fig. \ref{fig11} shows the corresponding histograms. Compared with the full model, we can observe that all the three variants achieve lower values across all three metrics, confirming the effectiveness of MMATT.

\begin{table}[htbp]
  % \footnotesize
  % \vspace{-0.5cm}
  \begin{center}
    \caption{Recognition results of the ablation experiment}
    \label{table7}
    \fontsize{8}{12}\selectfont
    \scalebox{0.99}{
    \begin{tabular}{cccc}
      \hline
      \textbf{Method} & \textbf{Overall Accuracy (\%)} & \textbf{F1-score (\%)} & \textbf{Average Accuracy (\%)} \\
      \hline
      \textbf{VMD-Fusion-CNN-MMATT-ImFL} & $\mathbf{91.68 \pm 1.00}$ & $\mathbf{90.04 \pm 0.76}$ & $\mathbf{89.87 \pm 0.63}$ \\
      \text{VMD-Fusion-CNN-MMATTA1-ImFL} & $89.93 \pm 0.94$ & $86.68 \pm 1.14$ & $84.58 \pm 1.44$ \\
      \text{VMD-Fusion-CNN-MMATTA2-ImFL} & $91.32 \pm 0.84$ & $89.02 \pm 0.46$ & $88.17 \pm 1.05$ \\
      \text{VMD-Fusion-CNN-MMATTA3-ImFL} & $91.16 \pm 1.22$ & $88.81 \pm 1.04$ & $89.03 \pm 0.70$ \\
      \hline      
      \text{VMD-Fusion-CNN-MMATTS1-ImFL} & $90.19 \pm 0.89$ & $87.09 \pm 0.98$ & $85.63 \pm 1.50$ \\
      \text{VMD-Fusion-CNN-MMATTS2-ImFL} & $91.49 \pm 0.73$ & $89.47 \pm 0.62$ & $89.16 \pm 0.51$ \\
      \hline
    \end{tabular}
    }
  \end{center}
  % \vspace{-0.5cm}
\end{table}

Additionally, to investigate the effect of attention block placement, we swap \textbf{Attention Block 1 and Block 2} in VMD-Fusion-CNN-MMATTS1-ImFL, and \textbf{Attention Block 2 and Block 3} in VMD-Fusion-CNN-MMATTS2-ImFL. 

The results demonstrate that all variants with swapped attention blocks achieve lower performance than the original model, indicating that the ordering of attention blocks significantly impacts performance. The results suggest that different network depths benefit from different attention mechanisms because shallow layers focus on local details, middle layers require multi-scale context, and deep layers depend on channel-wise feature recalibration. Our proposed structure, which assigns the most suitable attention type to each depth, therefore proves to be the most effective and well-justified.

\subsection{Analysis of Adjustable Class-Balanced Focal Loss}
\label{subsec11}

\begin{figure}[htbp]
  \vspace{-0.1cm}
  \centering
  \includegraphics[width=0.6\textwidth]{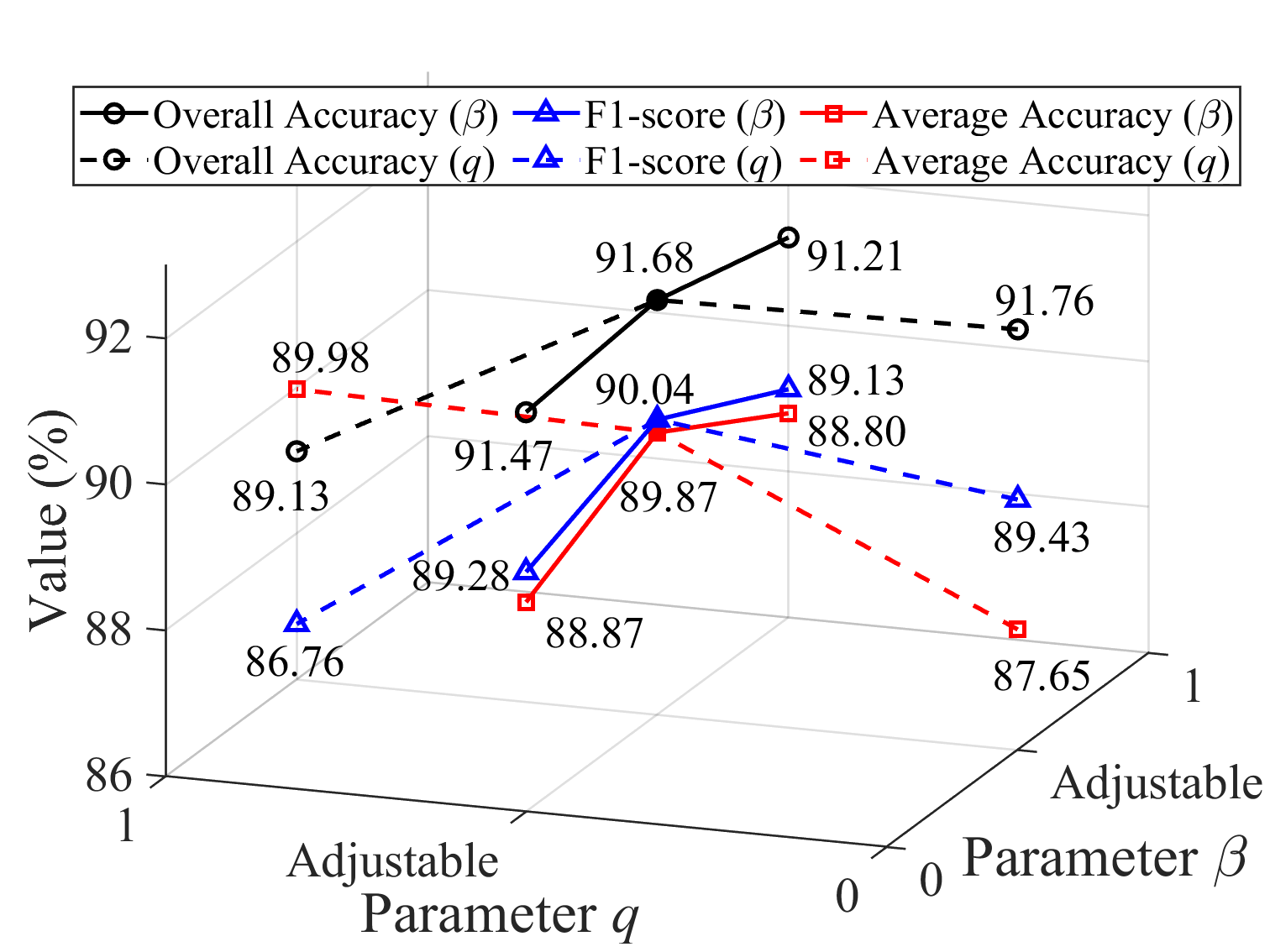} 
  \caption{The results of adjustable parameters and fixed parameters}
  \label{fig12}
\end{figure}

The tunability of ACBFL primarily stems from two adjustable parameters, $\beta$ and $q$. We investigate their influence on model recognition performance. The parameters $\beta$ and $q$ range from 0 to 1 and are determined via 5-fold cross-validation on the training set, with optimal values varying across training sets. We compare the model with adjustable $\beta$ and $q$ against those with fixed values ($0$ or $1$). Figure \ref{fig12} presents the results.

The results demonstrate that the model with adjustable $\beta$ and $q$ achieves the best performance, outperforming the models with $\beta = 0$ and $\beta = 1$ by $0.21\%$, $0.76\%$, $1.00\%$ and $0.47\%$, $0.91\%$, $1.07\%$ in terms of OA, F1-score, and AA, respectively. Compared with the model with $q = 0$, our model achieves a comparable OA but improves the F1-score by $0.61\%$ and AA by $2.22\%$; compared with the model with $q = 1$, it achieves a comparable AA while improving OA by $2.55\%$ and F1-score by $3.28\%$. These results validate the effectiveness of the proposed ACBFL.

\section{Conclusion}
\label{sec5}
This paper has developed an attention-based 1-D CNN for underwater acoustic target recognition. A novel 2-D DEMON spectral feature extraction and fusion scheme based on VMD and the 3/2-D spectrum has been proposed to enhance the representation of envelope information in ship-radiated noise. To improve recognition performance, we have designed the MMATT structure and have proposed two novel attention mechanisms: R-CISAM and MS-SFSAM. Furthermore, the class imbalance of ship-radiated noise data has been evaluated. Since long-tailed training data severely degrade recognition performance, we have proposed an Adjustable Class-Balanced Focal Loss. Extensive experiments on real-world ship-radiated noise data have been conducted to demonstrate the effectiveness of our proposed solutions.

\section*{Acknowledgments}
This study was supported by the National Natural Science Foundation of China under Grants 62192711 and 62371447.

\appendix
\section{Parameter Selection for the VMD Algorithm}
\label{app1}
The core parameters of the VMD algorithm are the number of modes $K$ and the penalty factor $\alpha$. We employ a comprehensive evaluation strategy based on three indicators: reconstruction error (RMSE), modal kurtosis, and central frequency overlap ratio. The mathematical definitions of these indicators are as follows:

\begin{enumerate}
	\item 
The reconstruction error measures the fidelity of the decomposition to the original information, which can be formulated as
\begin{equation}\label{eq16}
\mathrm{RMSE} = \sqrt{\frac{1}{N}\sum_{i=1}^{N}\bigl(x(t_i)-\hat{x}(t_i)\bigr)^2}
\end{equation}

	\item 
Kurtosis reflects the intensity of impulsive components in the signal. For each mode $u_{k}(t)$, its kurtosis is defined as
\begin{equation}\label{eq17}
\mathrm{Kurt}(u_k) = \frac{\frac{1}{N}\sum_{i=1}^{N}\bigl(u_k(t_i)-\bar{u}_k\bigr)^4}
{\left(\frac{1}{N}\sum_{i=1}^{N}\bigl(u_k(t_i)-\bar{u}_k\bigr)^2\right)^2} - 3
\end{equation}
where $\bar{u}_{k}$ denotes the mean of the mode, and the maximum kurtosis among all modes $\mathrm{Kurt}_{\max} = \max_{k} \mathrm{Kurt}(u_k)$ is invoked as the evaluation metric.

	\item 
The central frequency overlap ratio quantifies the overall separation of modes in the frequency domain. First, the central frequencies of the modes are sorted in ascending order as $f_{(1)}< f_{(2)} < \cdots < f_{(K)}$, and the normalized distance between a pair of modes is defined as
\begin{equation}\label{eq18}
d_{ij} = \frac{|f_{(i)} - f_{(j)}|}{B}
\end{equation}
where $B$ denotes the bandwidth. The central frequency overlap ratio (CFOR) is defined as the proportion of mode pairs whose normalized distance falls below a threshold $\delta$:
\begin{equation}\label{eq19}
\mathrm{CFOR} = \frac{2}{K(K-1)} \sum_{i=1}^{K-1} \sum_{j=i+1}^{K} \mathbf{1}_{\left\{ d_{ij} < \delta \right\}}
\end{equation}
where $\delta$ is a preset distance threshold (in this study, $\delta=0.1$). A smaller CFOR indicates a more uniform distribution of central frequencies across modes, better separation of frequency bands, and lower overall mode mixing.
\end{enumerate}

\begin{figure}[htbp] 
  \vspace{-0.1cm}
  \centering
  \begin{subfigure}[b]{.45\linewidth}
    \centering
    \includegraphics[width=\linewidth]{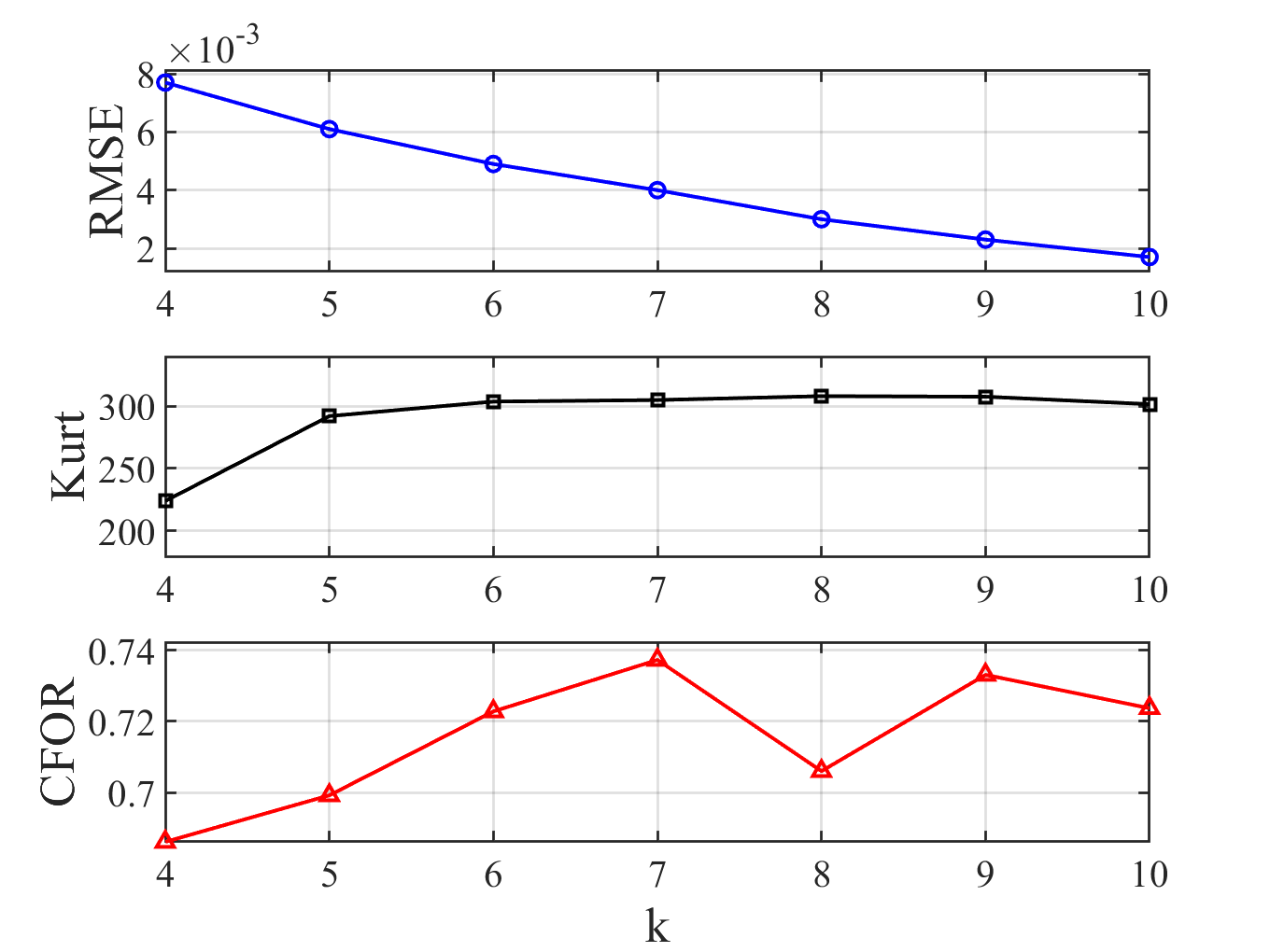}
    \caption{Three metrics versus $K (\alpha = 1000)$}
    \label{subfig10}
  \end{subfigure}
  \begin{subfigure}[b]{.45\linewidth}
    \centering
    \includegraphics[width=\linewidth]{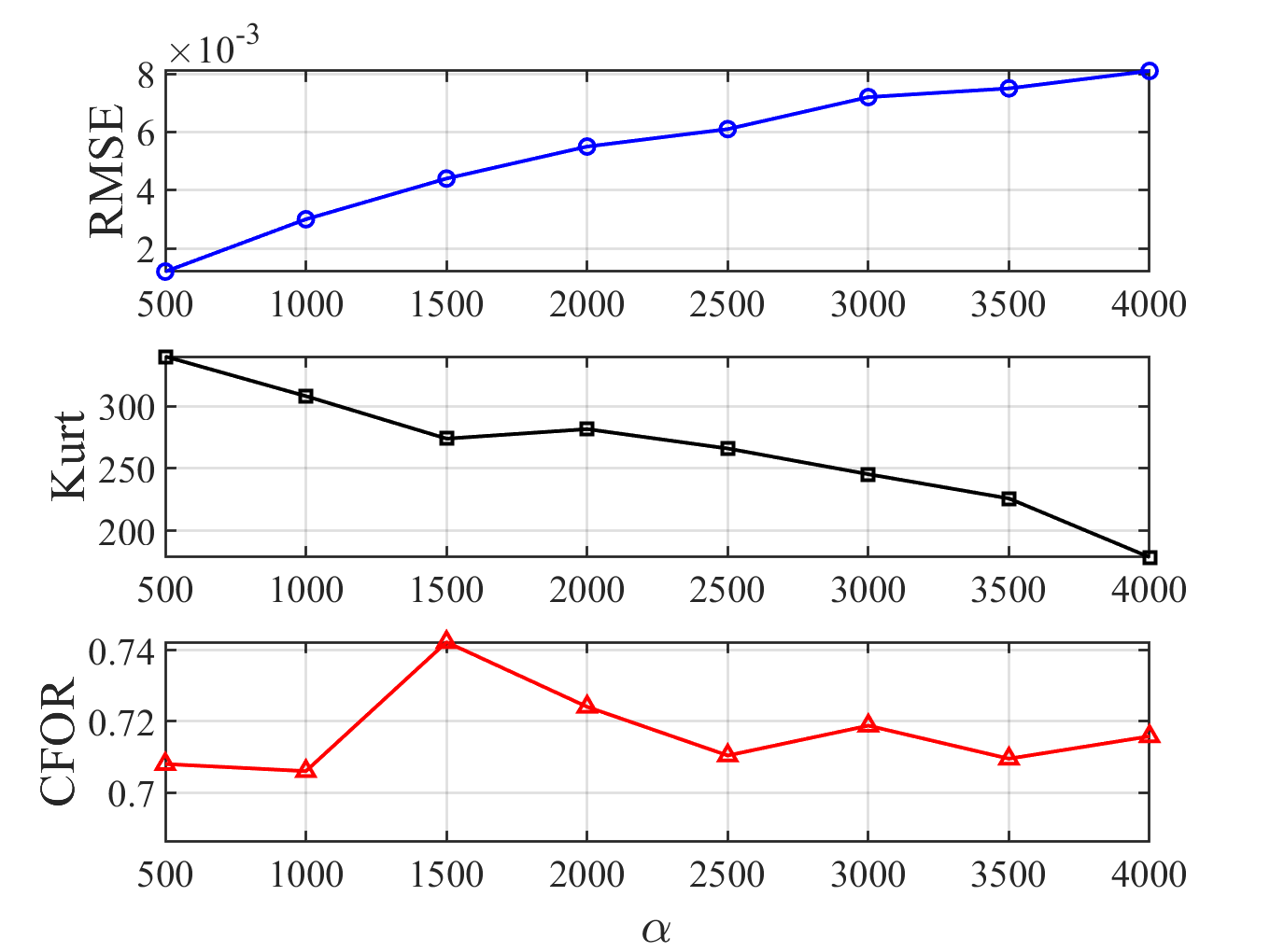}
    \caption{Three metrics versus $\alpha (K = 8)$}
    \label{subfig11}
  \end{subfigure} 
  \caption{Three evaluation metrics versus VMD parameters $K$ and $\alpha$}
  \label{fig13}
\end{figure}

In the experiments, we randomly select 10 samples from each vessel category as representatives and determine the values of $K$ and $\alpha$ by combining prior knowledge with quantitative metrics. Finally, we adopt $K = 8$ and $\alpha = 1000$. Fig. \ref{subfig10} shows the variations of the three indicators with different $K$ values under $\alpha = 1000$, and Fig. \ref{subfig11} shows the variations with different $\alpha$ values under $K = 8$. Experimental results demonstrate that $K = 8$ and $\alpha=1000$ constitute the optimal parameter combination when all metrics are considered comprehensively.

\section{Feature Comparison}
\label{app2}
This study focuses on extracting and utilizing modulation information from ship-radiated noise. Compared with other commonly used features such as time-frequency spectrogram, modulation features typically have a smaller data volume, which reduces the computational cost of subsequent processing and recognition. However, the modulation features only capture part of the information contained in ship-radiated noise, and the recognition performance of traditional modulation feature extraction methods is relatively limited in certain cases. 

\begin{figure}[htbp]
  \vspace{-0.1cm}
  \centering
  \includegraphics[width=0.6\textwidth]{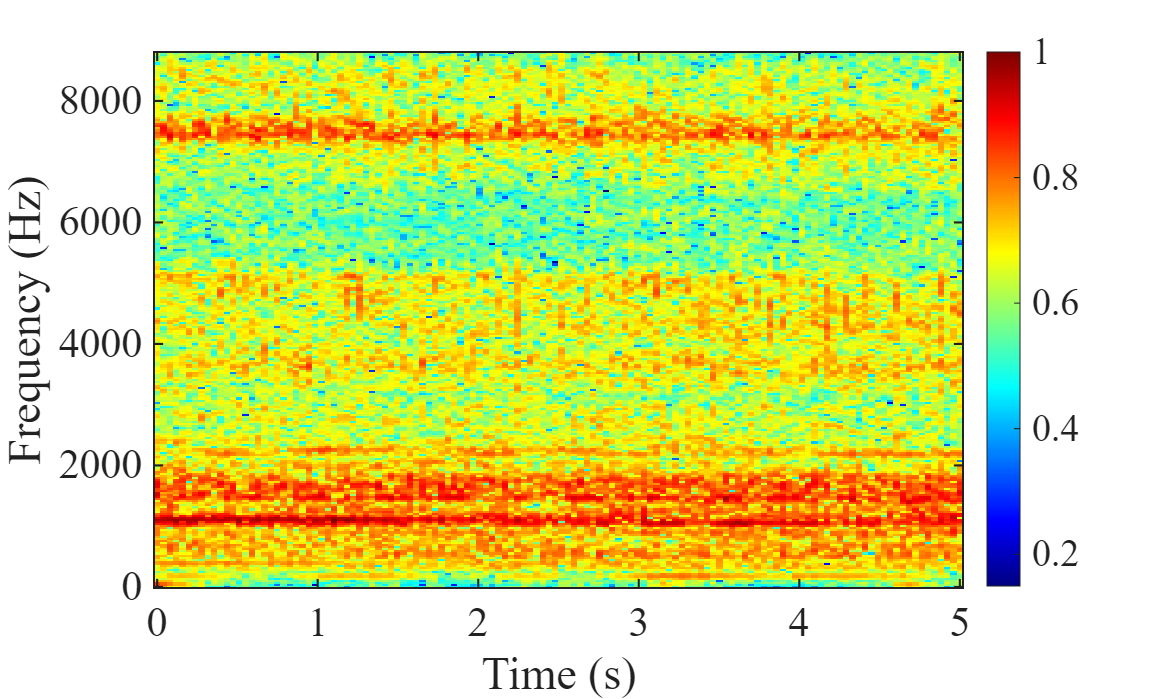}
  \caption{Time-frequency spectrogram}
  \label{fig14}
\end{figure}

Therefore, we compare the recognition performance of methods using both time-frequency spectrogram and DEMON spectrum. The time-frequency spectrogram has a size of $128 \times 256$, and an example is shown in Fig. \ref{fig14}. A 2-D CNN is applied to the time-frequency spectrogram and trained with the cross-entropy loss. Table \ref{table8} presents the experimental results, where TF-CNN denotes the model using time-frequency spectrogram.

\begin{table}[htbp]
  % \footnotesize
  % \vspace{-0.1cm}
  \begin{center}
    \caption{Recognition performance of different features}
    \label{table8}
    \fontsize{8}{12}\selectfont
    \scalebox{0.99}{
    \begin{tabular}{cccc}
      \hline
      \textbf{Method} & \textbf{Overall Accuracy (\%)} & \textbf{F1-score (\%)} & \textbf{Average Accuracy (\%)} \\
      \hline
      \text{TF-CNN} & $89.57 \pm 0.83$ & $82.82 \pm 1.59$ & $81.13 \pm 1.91$ \\
      \text{Trad-CNN} & $73.21 \pm 1.45$ & $63.79 \pm 1.82$ & $60.38 \pm 1.50$ \\
      \text{Filter-CNN} & $85.07 \pm 1.25$ & $80.70 \pm 2.40$ & $76.98 \pm 2.41$ \\
      \text{VMD-Fusion-CNN} & $89.50 \pm 0.72$ & $84.90 \pm 1.08$ & $82.58 \pm 1.75$ \\
      \hline
    \end{tabular}
    }
  \end{center}
  % \vspace{-0.5cm}
\end{table}

The results show that the time-frequency spectrogram outperforms traditional and bandpass-filtered DEMON spectrum, whereas our enhanced DEMON spectrum surpasses the time-frequency spectrogram in terms of F1-score and AA.

%%       \cite{<label>} ==> [1]
%%
%% Example citation, See 

\bibliographystyle{elsarticle-num}
\small
\bibliography{my_reference}



\end{sloppypar}
\end{document}